\documentclass[printer]{aa}
\usepackage{natbib}
\usepackage{amssymb}
\usepackage{graphicx}
\usepackage{hyperref}
\usepackage{comment}
\usepackage[caption=false]{subfig}
\graphicspath{{./figures/}}

\def\der{\hbox{d}}
\newcommand{\rhoism}{\mbox{$\rho_{\rm ISM} $}}
\newcommand{\vism}{\mbox{$v_{\rm ISM}$}}

\newcommand{\Mdisc}{\mbox{$M_\mathrm{disc}$}}

\newcommand{\Rin}{\ensuremath{R_\mathrm{in}}}
\newcommand{\Rout}{\ensuremath{R_\mathrm{out}}}

\newcommand{\Msun}{\mbox{$\mathrm{M}_{\odot}$}}

\newcommand{\yin}{\ensuremath{y_\mathrm{in}}}
\newcommand{\yout}{\ensuremath{y_\mathrm{out}}}

\makeatletter
\newcommand\footnoteref[1]{\protected@xdef\@thefnmark{\ref{#1}}\@footnotemark}
\makeatother

\begin{document}

\title{Characterising face-on accretion onto and the subsequent contraction of protoplanetary discs}
\titlerunning{Characterizing protoplanetary disc accretion and contraction}
\author{T.P.G. Wijnen\inst{1,2}, O.R. Pols\inst{1}, F.I. Pelupessy\inst{2,3}, S. Portegies Zwart\inst{2}}
\authorrunning{T.P.G. Wijnen et al.}
\institute{Department of Astrophysics/IMAPP, Radboud University Nijmegen, P.O. Box 9010, 6500 GL Nijmegen, The Netherlands\\
\email{thomas.wijnen@astro.ru.nl}
\and Leiden Observatory, Leiden University, PO Box 9513, 2300 RA Leiden, The Netherlands
\and Institute for Marine and Atmospheric research Utrecht, Utrecht University,
Princetonplein 5, 3584 CC Utrecht, The Netherlands
\offprints{T.P.G. Wijnen}
}
\date{Received ..../ Accepted ....}

\abstract{Observations indicate that stars generally lose their protoplanetary discs on a timescale of about 5 Myr. Which mechanisms are responsible for the disc dissipation is still debated.}
{Here we investigate the movement through an ambient medium as a possible cause of disc dispersal. The ram pressure exerted by the flow can truncate the disc and the accretion of material with no azimuthal angular momentum leads to further disc contraction.}
{We derive a theoretical model from accretion disc theory that describes the evolution of the disc radius, mass, and surface density profile as a function of the density and velocity of the ambient medium. We test our model by performing hydrodynamical simulations of a protoplanetary disc embedded in a flow with different velocities and densities.}
{We find that our model gives an adequate description of the evolution of the disc radius and accretion rate onto the disc. The total disc mass in the simulations follows the theoretically expected trend, except at the lowest density where our simulated discs lose mass owing to continuous stripping. This stripping may be a numerical rather than a physical effect. Some quantitative differences exist between the model predictions and the simulations. These are at least partly caused by numerical viscous effects in the disc and depend on the resolution of the simulation.}
{Our model can be used as a conservative estimate for the process of face-on accretion onto protoplanetary discs, as long as viscous processes in the disc can be neglected. The model predicts that in dense gaseous environments, discs can shrink substantially in size and can, in theory, sweep up an amount of gas of the order of their initial mass. This process could be relevant for planet formation in dense environments.
}

\maketitle

\section{Introduction}

Star formation generally occurs in clustered environments \citep{lada03}, meaning that the birth environment of stars is dense both in terms of stellar and gas densities. These conditions can influence the subsequent evolution of newborn stars and their protoplanetary discs via, for example photo-evaporation \citep[e.g.][]{balog07, guarcello07, guarcello09, fang12, facchini16}, close stellar encounters \citep[e.g.][]{breslau14, rosotti14, vincke15, portegies_zwart16}, and nearby supernovae \citep[e.g.][]{chevalier00, ouellette07, lichtenberg16}. 

The relative lifetimes of the different evolutionary stages of protoplanetary discs have not yet been determined with certainty \citep[see e.g.][ and references therein, who find e-folding times for the frequency of warm dust and gas discs of 6 and 4 Myr, respectively]{cloutier14}. It is still not fully understood which mechanism(s) dominate(s) the removal of gas from the disc during its evolution. Among the suggested mechanisms are photo-evaporation, angular momentum transport (e.g. by magnetorotational and gravitational instabilities), magnetic winds, and magnetic braking \citep[e.g.][ and references therein]{armitage11}. Another process that may influence the evolution of protoplanetary discs is their movement through an ambient medium. 

In a previous work \citep[][ Paper I hereafter]{wijnen16} we investigated to what extent stars surrounded by a protoplanetary disc can sweep up gas from their surroundings. We found that the motion of a star and its protoplanetary disc through a gaseous environment decreases the size of the disc due to two processes: (1) stripping of disc material by the ram pressure exerted by the interstellar medium (ISM) and (2) the accretion of ISM with little to no azimuthal angular momentum. The latter process lowers the specific angular momentum, i.e. the angular momentum per unit mass, of the gas in the disc and was also found by \citet{moeckel09}. These authors pointed out that this process causes the gas in the disc to follow a tighter orbit that corresponds to its new specific angular momentum. As disc material migrates inwards, the surface density profile of the protoplanetary disc increases at smaller radii. The influence of these two processes depends on, among other parameters, both the density and velocity of the ISM with respect to the disc. The effect of ram-pressure stripping and the redistribution of angular momentum owing to the accretion of ISM on the lifetimes of protoplanetary discs have received comparatively little attention.

The inward migration of gas in the disc and the increase of the surface density profile could play a role in planet formation and/or migration. \citet{ronco14} found that a disc with a steep surface density profile, i.e. more mass at smaller radii, is more likely to form a planet with a significant water content in the habitable zone. Likewise, `hot Jupiters', massive planets in a close orbit around their host star, are believed to have formed at larger radii and the inward migration of gas in the disc may have aided them in this process.

Here we present a theoretical model that describes the evolution of the mass, radius, and surface density profile of the disc when it is subject to accretion from a face-on ISM flow. We test this model by performing smoothed particle hydrodynamic (SPH) simulations of a protoplanetary disc embedded in a flow with different densities and velocities. As discussed in paper I (see Sect. 2 therein), the following (external) physical effects determine the size of the disc and therefore affect the process of entraining ISM by the protoplanetary disc: (1) ram pressure stripping, (2) redistribution of angular momentum, (3) close stellar encounters, and (4) photo-evaporation. We do not take the latter two processes into account in this work, but note that they have been explored by e.g. \citet{breslau14, rosotti14, portegies_zwart16} and \citet{balog07, guarcello07, guarcello09, fang12,facchini16}. In order for photo-evaporation to truncate discs to radii $<100$ AU on timescales of less than 10 Myr requires ultraviolet fluxes that are at least an order of magnitude higher than typical values \citep[see Figs. 3 and 4 of][ and references therein]{adams10}. Our model applies to the embedded phase of star formation in which the effect of photo-evaporation is expected to be reduced by the presence of the primordial gas. Simulations of close encounters have shown that an average stellar density of 500 pc$^{-3}$ (with a core density of $4 \times 10^4$ pc$^-3$) is required to truncate 40\% of the disc to radii $<$ 100 in 5 Myr \citep{vincke15}. Future work, which combines the four processes, can compare the relative importance of all these mechanisms on the survival of discs in star-forming regions.
The purpose of this paper is to present a theoretical model for the processes of ram pressure stripping and redistribution of angular momentum that can be implemented in future studies.

Our theoretical model is presented in Sect. \ref{sec:theory_param}. The set-up of our simulations is discussed in Sect. \ref{sec:setup_param} and we present the results in Sect. \ref{sec:results_param}. This is followed by a discussion (Sect. \ref{sec:dis_param}) and conclusion (Sect. \ref{sec:conclusion_param}).

\section{Theoretical framework}\label{sec:theory_param}

We found in paper I that if the ram pressure that is exerted by the flow is sufficient all disc material beyond a certain radius is stripped and dragged along with the flow. In Sect. \ref{sec:rtrunc_param} we give a theoretical derivation for this radius. The subsequent evolution of the disc is not dominated by ram pressure stripping but by the redistribution of angular momentum within the disc, as material with no azimuthal angular momentum is accreted. In Sect \ref{sec:timescales_param} we discuss the relevant timescales for our model and the solution to our model that we use to compare with the simulations is given in Sect. \ref{sec:model_param}.

\subsection{Ram pressure truncation radius}\label{sec:rtrunc_param}

As discussed in the introduction, the disc accretes material that has no azimuthal angular momentum. \cite{chevalier00} proposed a method to derive the radius beyond which disc material is stripped. \cite{chevalier00} equates the gravitational restoring force per unit surface area, i.e. `pressure', of the disc with the ram pressure exerted by the ISM. We used this derivation in paper I. However, we find that in our simulations the radius beyond which all disc material is stripped is smaller than the truncation radius derived by \cite{chevalier00}. We therefore follow a slightly different reasoning than \cite{chevalier00} and we find a consistent estimate of the disc truncation radius that differs by a small constant factor. 

We assume the gas in the disc moves in Keplerian orbits around the star. During its orbital motion, the ISM flow injects momentum into the gas perpendicular to its orbit, which slightly inclines the orbit with respect to the mid-plane of the disc. As the gas in the disc moves to the opposite side of the disc with respect to the star, it receives momentum from the ISM flow that cancels the inclination of its orbit. In other words, if the Keplerian timescale of the gas element is shorter than the timescale on which momentum is added, the flow is not able to inject enough momentum to strip the material from the disc. We approximate the timescale on which momentum is added as

\begin{equation}\label{eq:taup_param}
\tau_{\dot{p}} (r) = \frac{p(r)}{\dot{p}(r)}=\frac{m(r) v_{\rm kep}(r)}{\dot{m}(r) \vism}
,\end{equation}
where $p(r)$ is the momentum of a ring of gas in the disc at distance $r$ from the star, $\dot{p}(r)$ is the change in momentum of this ring, $m(r)$ the mass of the ring, $\dot{m}(r)$ its change in mass, $v_{\rm kep}(r)$ is the rotational velocity, and $\vism$ is the velocity of the ISM with respect to the disc. We equate this timescale to the Keplerian timescale, $\tau_{\rm kep}(r) = 2 \pi r / v_{\rm kep}(r)$, which gives

\begin{equation}\label{eq:Rtrunc_param}
R_{\rm trunc} = \left(\frac{GM_*\Sigma_0 r_0^n}{2 \pi \rhoism \vism^2}\right)^{\frac{1}{n+2}}
,\end{equation}
where $G$ is the gravitational constant, $M_*$ the mass of the star, $\rhoism$ the density of the ISM and we have assumed that we can write the surface density profile of the disc as $\Sigma(r) = \Sigma_0 (r/r_0)^{-n}$, where $r_0$ is an arbitrary radius to which the surface density profile is scaled. This approximation of the truncation radius differs by a factor $(2 \pi)^{-1/(n+2)}$ from the derivation by \citet{chevalier00}. This a factor $0.6$ for a typical value of $n=1.5$. Both derivations approximate the truncation radius to first order and are consistent with each other. However, we find that our derivation of the truncation radius agrees better with our simulations, so we use Eq. \ref{eq:Rtrunc_param} in the rest of this work. 

\subsection{Disc evolution in the presence of ISM accretion}\label{sec:redistribution_param}

As the disc accretes material that has no azimuthal angular momentum, the specific angular momentum in the disc decreases. Material in the disc therefore migrates to an orbit at a smaller radius that corresponds to its new specific angular momentum. Furthermore, the gas in the disc is generally subject to viscous forces. In the following we assume that the mass flux $\rhoism \vism$ is uniform in space and is fully accreted by the each surface element of the disc. Following the formalism for the evolution of a geometrically thin accretion disc in \citet{frank02}, assuming conservation of mass and angular momentum, we can derive a differential equation, which describes the radial motion of the gas at radius $r$ subject to these effects, as follows:
\begin{equation} \label{eq:dr-dt-viscous_param}
\frac{\der r}{\der t} = - 2\rhoism \vism \, \frac{r}{\Sigma} - \frac{3}{r^{1/2}\Sigma} \frac{\partial}{\partial r} (r^{1/2} \nu\Sigma),
\end{equation}
where $\Sigma$ is the surface density and $\nu$ the viscosity of the gas in the disc. Similarly, we can derive a differential equation for the evolution of the surface density profile,
\begin{equation} \label{eq:dSigma-dt-viscous_param}
\frac{\partial\Sigma}{\partial t} = 5 \rhoism \vism + \frac{3}{r} \frac{\partial}{\partial r} \Bigg[ r^{1/2} \frac{\partial}{\partial r} (r^{1/2} \nu\Sigma) \Bigg].
\end{equation}
For a derivation of these equations, see Appendix \ref{ap:disc_eqs_deriv}. Eqs. \ref{eq:dr-dt-viscous_param} and \ref{eq:dSigma-dt-viscous_param} are equivalent to Eqs. 5.8 and 5.9 in \citet{frank02}, with the addition of an ISM accretion term involving $\rhoism \vism$ in each equation. As expected, this term leads to overall contraction of the disc (cf. Eq. \ref{eq:dr-dt-viscous_param}) and to an increase of the surface density. Accretion alone would give a term equal to $\rhoism \vism$ in Eq.~\ref{eq:dSigma-dt-viscous_param}; the enhancement by a factor of 5 is the result of the overall contraction of the disc. Eq. \ref{eq:dSigma-dt-viscous_param} has the form of a diffusion equation with a source term $5 \rhoism \vism$. The general solution of this equation requires numerical methods. An added difficulty is that the disc viscosity $\nu$ is an unknown function of the physical parameters, which is a well-known problem in disc physics. In the next section we discuss to what extent we can neglect the viscous evolution of the disc.

\subsection{Timescales}\label{sec:timescales_param}

We limit ourselves to analysing the circumstances under which the contributions from ISM accretion and from viscous torques dominate the evolution of the disc. For this purpose, we assume -- as is usual in the thin disc approximation -- that the scale height $H$ of the disc at a certain radius is given by
\begin{equation} \label{eq:disc-thickness_param}
H \approx c_s \sqrt{\frac{r}{GM_{*}}}\, r = \frac{c_s}{\Omega},
\end{equation}
where $c_s$ is the local sound speed and $\Omega = \sqrt{GM_{*}/r^3}$ the angular velocity. Furthermore we assume the viscosity is given by the \citet{shakura73} $\alpha$-prescription,
\begin{equation} \label{eq:alpha-visc_param}
\nu = \alpha\, H c_s = \alpha\, \frac{c_s^2}{\Omega}.
\end{equation}
The two terms in Eq. \ref{eq:dSigma-dt-viscous_param} affect the surface density on different timescales. The local timescale for viscous evolution is, on dimensional grounds,
\begin{equation} \label{eq:tauvisc_param}
\tau_{\nu}(r) \approx \frac{r^2}{3\nu} \approx \frac{\sqrt{GM_{*}r}}{3\alpha c_s^2}.
\end{equation}
The local timescale for accretion or mass loading from the ISM is
\begin{equation}\label{eq:taum_param}
\tau_{\dot{m}} (r) = \frac{\Sigma(r)}{5\rhoism \vism},
\end{equation}
which can be seen as the timescale on which a disc annulus accretes of the order of its own mass. The evolution of the disc at a certain radius is dominated by the process with the shortest timescale. Since in most realistic situations both $\Sigma$ and $c_s$ decrease towards larger radius (the latter may be constant in an isothermal disc), $\tau_{\nu}$ typically increases outwards and $\tau_{\dot{m}}$ decreases outwards. We can thus distinguish three possible situations:

\begin{enumerate}
\item $\tau_{\dot{m}} \gg \tau_{\nu}$ everywhere in the disc, i.e. the accretion of ISM is so slow compared to viscous effects that it has a negligible effect on the evolution of the disc.
\item $\tau_{\dot{m}} > \tau_{\nu}$ in the inner part of the disc, where the evolution of the disc is viscosity dominated, while in the outer parts $\tau_{\dot{m}} < \tau_{\nu}$ and the evolution of the disc is driven by the effects of accretion. 
\item $\tau_{\dot{m}} \ll \tau_{\nu}$ everywhere in the disc, in which case the accretion of ISM is rapid enough that viscous effects can be ignored.
\end{enumerate}

Furthermore, Eq. \ref{eq:dr-dt-viscous_param} shows that ISM accretion always causes disc matter to move inwards ($\der r/\der t < 0$), which is a direct result of mass loading without angular-momentum accretion. Viscous torques can cause either inward or outward motion, depending on the slope of the function $r^{1/2} \nu\Sigma$. For a disc described by the Shakura-Sunyaev formalism and in which the product $c_s^2 \Sigma(r)$ follows a power law, $c_s^2 \Sigma \propto r^{-p}$, we can write Eq.~\ref{eq:dr-dt-viscous_param} as
\begin{equation} \label{eq:dr-dt-timescales_param}
\frac{\der r}{\der t} = - \frac{2}{5}\, \frac{r}{\tau_{\dot{m}}} - (2-p)\, \frac{r}{\tau_{\nu}}.
\end{equation}
Thus, as long as $p<2$ the viscous torques also cause matter to spiral inwards, but for a (locally) steep surface density gradient with $p>2$ matter moves outwards as a result of viscosity. The latter can be expected to occur at the outer edge of the disc where $\Sigma(r)$  drops off sharply as an effect of the flow. For the special case $p=2,$ the gas in the disc is stationary in radius.

With the same assumptions Eq.~\ref{eq:dSigma-dt-viscous_param} can be written as
\begin{equation} \label{eq:dSigma-dt-timescales}
\frac{\partial\Sigma}{\partial t} = \frac{\Sigma}{\tau_{\dot{m}}} + (2-p)({\textstyle\frac{3}{2}}-p) \, \frac{\Sigma}{\tau_{\nu}}.
\end{equation}
The viscous torques thus have no effect on $\Sigma(r)$ when either $p=2$ or $p=\frac{3}{2}$. The first case corresponds to the stationary situation discussed above, while for $p=\frac{3}{2}$ gas moves inwards at all radii, but in such a way as to maintain the same surface density profile.

\begin{figure}[!ht]
\centering
    \includegraphics[width=0.49\textwidth]{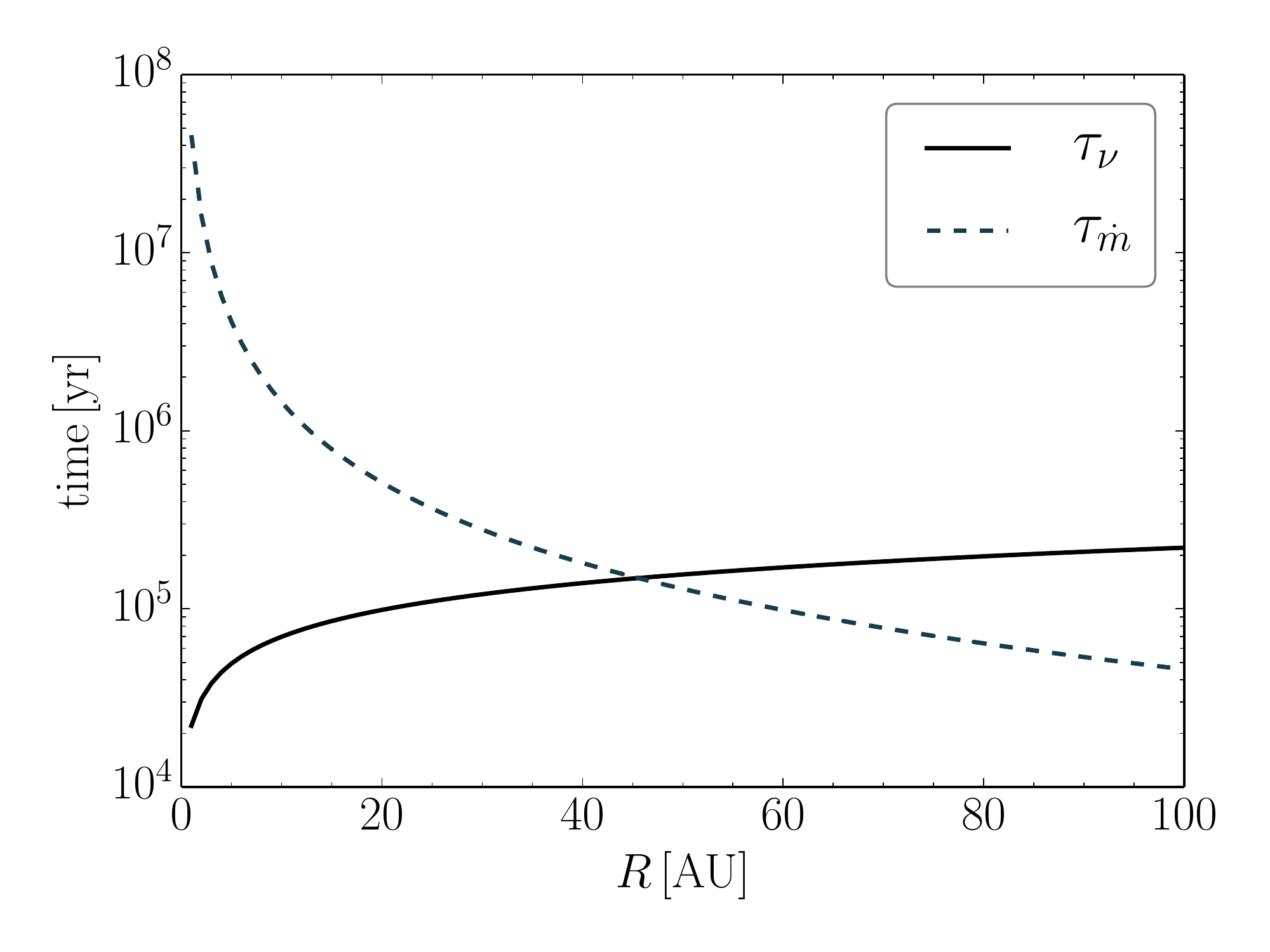}
    \caption{Different timescales as a function of the radius in the disc, using the disc parameters of one of our simulations, $\vism = 3\,$km/s and $\rhoism = 1.9 \times 10^{-19}$g/cm$^3$. This is the case with the longest accretion timescale, $\tau_{\dot{m}}$, which is not in the regime where Bondi-Hoyle accretion dominates. The dashed line is the timescale on which mass is loaded on the disc (see Eq. \ref{eq:taum_param}) and the solid line represents the viscous timescale (see Eq. \ref{eq:tauvisc_param}. \label{fig:timescales_param}}  
\end{figure}

As an example, we have estimated both timescales in Fig. \ref{fig:timescales_param}, assuming that $\vism = 3\,$km/s and $\rhoism = 1.9 \times 10^{-19}$g/cm$^3$. In accretion disc theory, $\alpha \approx 0.01$ \citep{armitage11}. Since in our hydrodynamical simulations (Sect.~\ref{sec:setup_param}) we assume $\alpha_{\rm SPH} = 0.1$, which corresponds to a physical $\alpha$ of roughly 0.02 (following \citep{artymowicz94}), we decided to use the latter value. 
This case corresponds to our simulation with the lowest mass flux and the longest $\tau_{\dot{m}}$, which is still outside the regime where Bondi-Hoyle accretion is expected to dominate the accretion process (Sect.~\ref{sec:BHaccretion_param}.) and the disc evolution is no longer properly described by Eqs. \ref{eq:dr-dt-viscous_param} and \ref{eq:dSigma-dt-viscous_param} (see Sect. \ref{sec:BHaccretion_param}). The case in Fig. \ref{fig:timescales_param} corresponds to regime 2 because ISM mass loading dominates the evolution of the outer parts of the disc.
Therefore, in this case and in cases with a higher mass flux and shorter $\tau_{\dot{m}}$, the actual size of the disc is determined by the accretion of ISM and not by viscous effects. We may neglect viscous effects in the outer parts of the disc on timescales shorter than
\begin{equation}\label{eq:tauvisc-dom_param}
 \tau_{\nu,\,\mathrm{dom}}=  {\tau_{\dot{m}}}^{1/6}(R_0) \, {\tau_{\nu}}^{5/6}(R_0);
\end{equation}
see Appendix \ref{app:viscous_effects_param}. Here $R_0$ is the initial outer radius of the disc, after accounting for possible ram pressure stripping. On longer timescales viscous effects dominate the evolution of the disc. This timescale is $1.7 \times 10^5$ yr for the parameters shown in Fig. \ref{fig:timescales_param}.
The weak dependence of $\tau_{\nu,\,\mathrm{dom}}$ on the accretion timescale means that in our simulation with the highest mass flux ($\vism = 30\,$km/s and $\rhoism = 1.9 \times 10^{-17}$g/cm$^3$) $\tau_{\nu,\,\mathrm{dom}}$ is only moderately shorter, i.e. $4\times 10^4$ yr.  In all cases this is well beyond the maximum simulation time of $10^4$ year (Sect.~\ref{sec:setup_param}).

\subsection{Analytical model for inviscid disc evolution}\label{sec:model_param}

Since we can neglect viscous effects at least during the time span of our simulations, we can simplify Eqs. \ref{eq:dr-dt-viscous_param} and \ref{eq:dSigma-dt-viscous_param} to describe the accretion and contraction of discs in the inviscid case. Neglecting the viscosity terms results in the following equations:
\begin{equation} \label{eq:drdt_param}
\frac{\der r}{\der t} = - 2\rhoism \vism \, \frac{r}{\Sigma(r, t)}
\end{equation}
and
\begin{equation} \label{eq:dsigmadt_param}
\frac{\partial \Sigma}{\partial t} = 5 \rhoism \vism.
\end{equation}
The simple form of Eq.~\ref{eq:dsigmadt_param} allows us to write the solution as
\begin{equation}\label{eq:sigmari_param}
\Sigma(r,t) = \Sigma_0(r) + 5 \int_0^t \rhoism \vism \,\der t',
\end{equation}
where $\Sigma_0(r)$ is the initial surface density at radius $r$. If $\rhoism \vism$ is a known function of time, this can be integrated directly; in that case it is useful to define a dimensionless parameter $\tau$ as follows:
\begin{equation} \label{eq:tau-integral}
\tau = \frac{5}{\Sigma_0(r_0)} \int_0^t \rhoism \vism \,\der t',
\end{equation}
such that $\der\tau/\der t = 5\rhoism \vism/\Sigma_0(r_0)$; $r_0$ is an arbitrary but constant reference radius that initially lies within the disc.\ The parameter $\tau$ thus increases linearly with the total accreted mass flux, and monotonically -- although in general non-linearly -- with time. If we further assume a power-law shape for the initial surface density distribution, i.e.\ $\Sigma_0(r) = \Sigma_0\,(r/r_0)^{-n}$, where for brevity we write $\Sigma_0 \equiv \Sigma_0(r_0),$ and we define a dimensionless radius $y = r/r_0$, then Eq.~\ref{eq:drdt_param} can be written as
\begin{equation} \label{eq:dydtau}
\frac{\der y}{\der\tau} = - \frac{2}{5} \, \frac{y}{y^{-n} + \tau}.
\end{equation}
The solution to Eq.~\ref{eq:dydtau} is scale-free and only depends on the power-law exponent $n$. The solution $y(\tau)$ can be scaled according to the physical parameters $r_0$, $\Sigma_0$ (or equivalently, the initial disc mass) and the integrated mass flux $\int \rhoism \vism\, \der t$. Eq.~\ref{eq:dydtau} describes the movement of all annuli in the disc, but in practice it only needs to be solved for the inner and outer disc radii, \yin\ and \yout.

The evolution of the surface density follows from the solution $y(\tau)$ and Eq.~\ref{eq:sigmari_param},
\begin{equation} \label{eq:sigma-tau}
\Sigma(r,t) = \Sigma_0\, ( y^{-n} + \tau ).
\end{equation}
The mass of the disc then follows from the surface density distribution as
\begin{eqnarray}\label{eq:Mdisc_param}
&& M_{\rm disc} (t) =  \int_{\Rin}^{\Rout} 2\pi r\, \Sigma(r,t) \,\der r \nonumber\\
&& =  2\pi r_0^2 \Sigma_0 \bigg[ \frac{1}{2-n} (\yout^{2-n} - \yin^{2-n}) + \frac{1}{2} (\yout^2 - \yin^2) \tau \bigg],
\end{eqnarray}
with $\Rin(t) = \yin(\tau) \, r_0$ and $\Rout(t) = \yout(\tau) \, r_0$. Eq.~\ref{eq:Mdisc_param} also gives the initial disc mass $M_\mathrm{disc,0}$ by setting $\tau=0$, which in turn defines $\Sigma_0$.
The time derivative of Eq.~\ref{eq:Mdisc_param}, making use of Eqs.~\ref{eq:tau-integral} and \ref{eq:dydtau}, is
\begin{eqnarray} \label{eq:dMdisk-dt}
\dot{M}_{\rm disc}(t) & = & 2\pi r_0^2 \Sigma_0 \, \frac{(\yout^2 - \yin^2)}{10} \, \frac{\der\tau}{\der t} \nonumber \\
& = & \rhoism \vism \, \pi (\Rout^2 - \Rin^2).
\end{eqnarray}
The latter expression equals the ISM accretion rate, $\dot{M}_{\rm ISM}$, which is expected from the geometric cross-section of the disc. Using Eq. \ref{eq:Mdisc_param}, the amount of accreted ISM, $\Delta M_{\rm ISM}$, can be expressed as
\begin{equation}\label{eq:deltaM_param}
\Delta M_{\rm ISM}(t) = M_{\rm disc}(t) - M_{\rm disc,0}.
\end{equation}

\begin{figure*}[!ht]
\centerline{%
\includegraphics[width=0.48\textwidth]{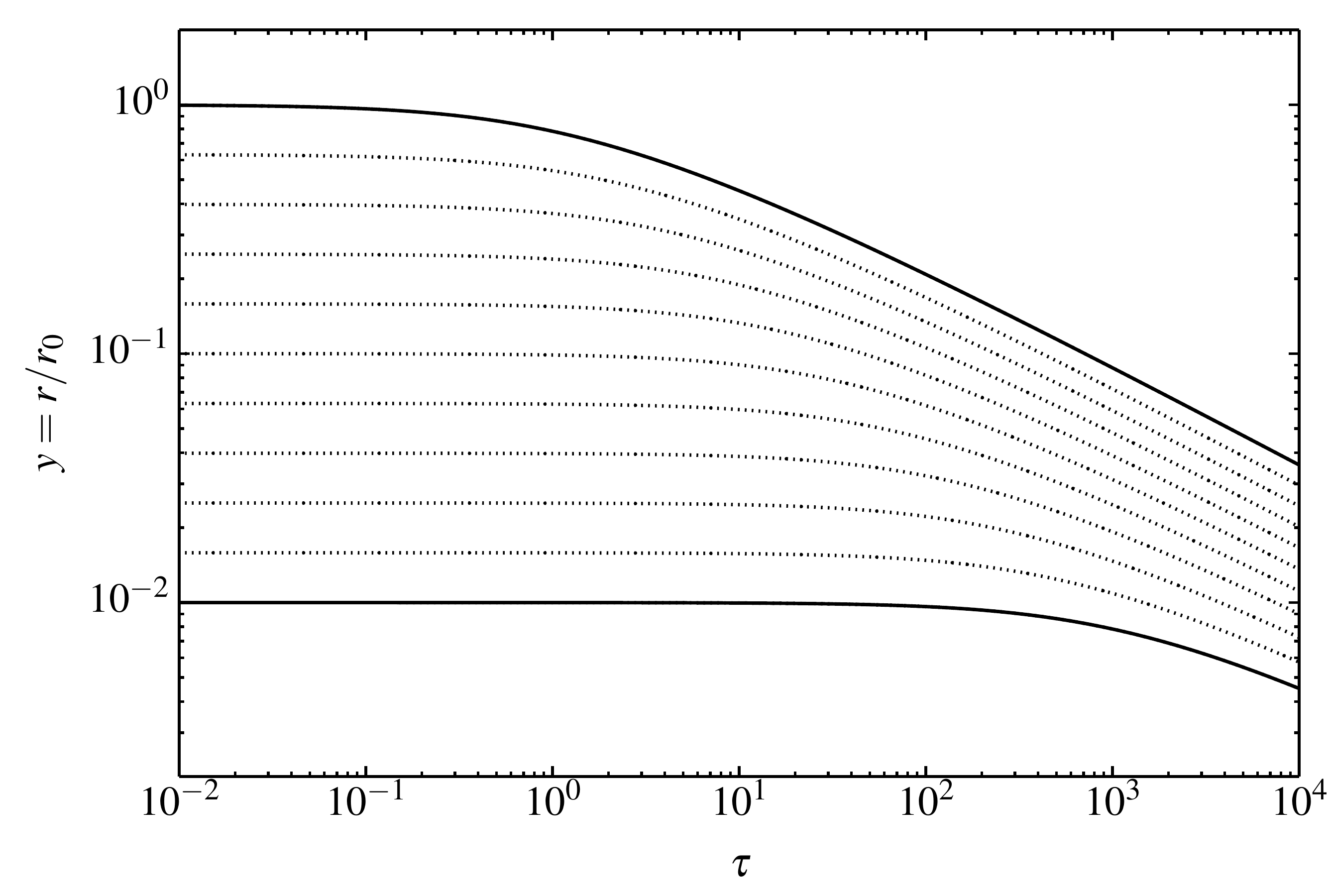}\hfill
\includegraphics[width=0.48\textwidth]{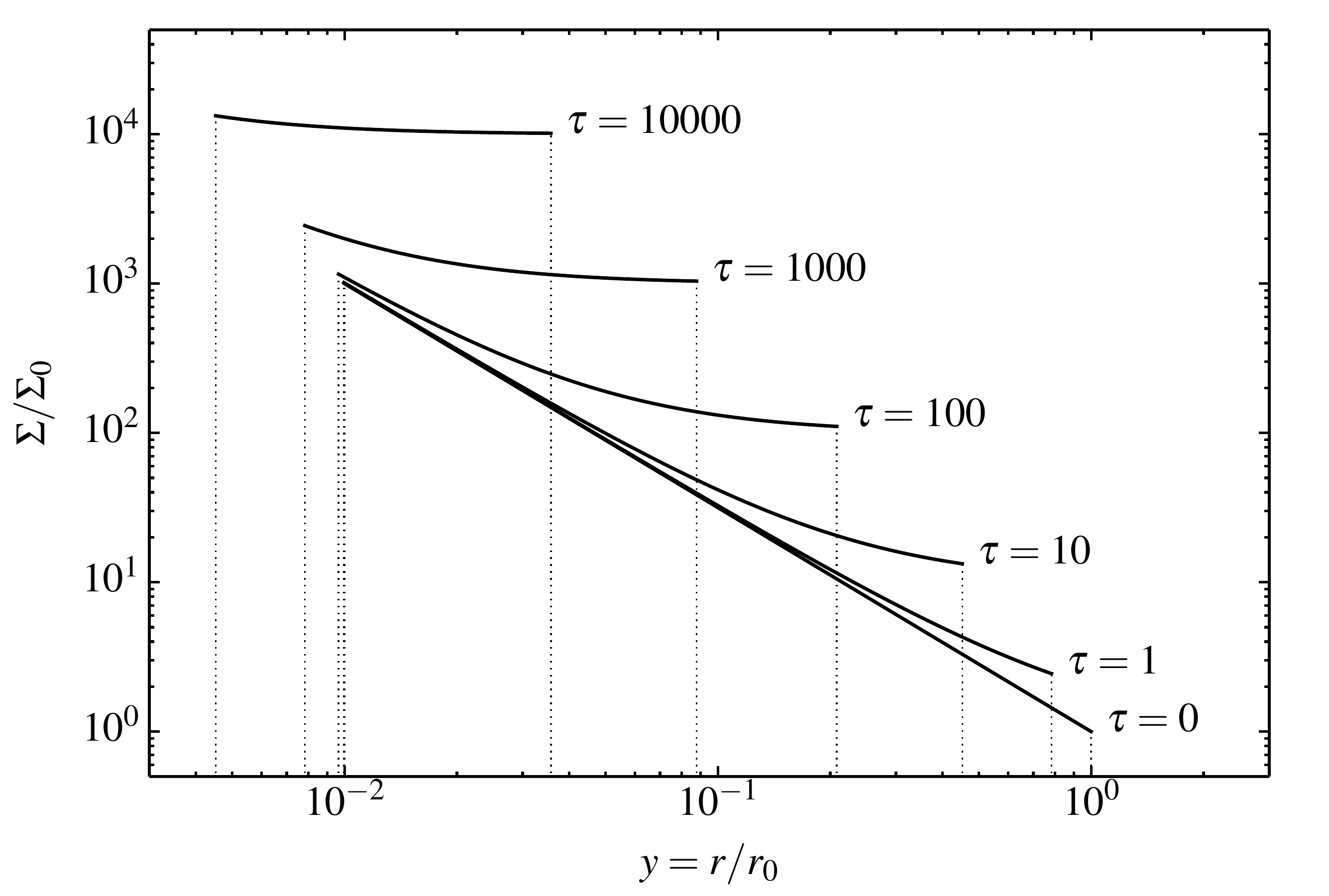}}
\caption[]{Left panel: Solutions to Eq.~\ref{eq:dydtau} for a disc with $n = 1.5$, initially extending from $r = 0.01 r_0$ to $r = r_0$. Solid lines show the evolution of the inner and outer disc radius and dotted lines show several intermediate radii.  Right panel: Corresponding surface density distribution $\Sigma(r,t)/\Sigma_0$ from Eq.~\ref{eq:sigma-tau} for several values of $\tau$.}
\label{fig:solution}
\end{figure*}

Eq.~\ref{eq:sigma-tau} implies that the accreted gas dominates over the initial surface density for $\tau > y^{-n}$ for an annulus located at radius $y$. For a time-independent mass flux this happens at time $t > \tau_{\dot{m}}(r)$. From Eq.~\ref{eq:dydtau} it follows that on this timescale the solution approaches a power law, $y \propto \tau^{-2/5}$, independent of $n$. This happens earlier for larger $y_0$; that is, the outer parts of the disc start contracting sooner than the inner parts (see Fig.~\ref{fig:solution}). This is accompanied by a flattening of the surface-density profile, as follows from Eq.~\ref{eq:sigma-tau}: for $\tau > {\yin}^{-n}$, $\sigma \approx \tau$ independent of radius (see the right-hand panel of Fig.~\ref{fig:solution}). Eq.~\ref{eq:Mdisc_param} shows that for $\tau > {\yout}^{-n}$ the second term inside the square brackets dominates and thus $M_{\rm disc} \propto \tau^{1/5}$, since $\yout \propto \tau^{-2/5}$. In other words, on long timescales the disc mass grows much more slowly than linearly with time. We need to keep in mind, however, that viscous effects can no longer be ignored over very long timescales (Sect.~\ref{sec:timescales_param}).

We use this model to predict the evolution of the protoplanetary disc in our simulations. We assume a constant density and velocity as a function of time, but this does not provide an analytical solution to Eq. \ref{eq:dydtau}, which has to be integrated numerically. The radius at which we start the integration is determined by the ram pressure stripping. In our simulations the disc has an initial radius, $R_{\rm disc}$, of 100 AU, but depending on the density and velocity of the flow ram pressure may almost instantaneously remove disc material beyond a certain radius. We therefore take the minimum of $R_{\rm disc}$ and $R_{\rm trunc}$, as determined in Sect. \ref{sec:rtrunc_param}, as the initial condition for the integration.

\subsection{Bondi-Hoyle accretion}\label{sec:BHaccretion_param}
 
At low velocities the Bondi-Hoyle accretion radius becomes comparable to the initial radius we assume for the protoplanetary disc. The Bondi-Hoyle accretion radius, $R_{\rm BH}$, is given by \citep{Bondi52, shima85},\textbackslash
\begin{equation}\label{eq:R_BH_param}
R_{\rm BH} = \frac{2GM_{*}}{c_s^2 +\vism^2}.
\end{equation}
The resulting accretion rate is
\begin{equation}\label{eq:mdotBH_param}
\dot{M}_{\rm BH} = \pi R_{\rm BH}^2 \rhoism \sqrt{\lambda^2 c_s^2 + \vism^2},
\end{equation}
where $\lambda$ is a constant of order unity that depends on the equation of state of the gas. In the isothermal case, which we assume, $\lambda = e^{3/2}/4$.  

In our simulations $R_{\rm BH}$ is 78 AU for $\vism =3\,\mathrm{km/s}$ and 651 AU for  $\vism =1\,\mathrm{km/s}$. Since we assume a disc radius of 100 AU, the effective cross section of the star and protoplanetary disc system is larger than its physical size when the velocity of the ISM is $\lesssim 2.7$ km/s. We therefore expect that more ISM is accreted than is estimated from the geometric cross-section. Although most of this additional accretion occurs via the accretion wake onto the star, some material from outside the disc radius is gravitationally focussed towards the disc and falls onto its surface. If more material with no azimuthal angular momentum is accreted by the disc than expected from Eq.~\ref{eq:dMdisk-dt}, the shrinking process of the disc is sped up.

We can define a typical timescale for Bondi-Hoyle accretion as
\begin{equation}\label{eq:tau_BH_param}
\tau_{\rm BH} = \frac{R_{\rm BH}}{\sqrt{c_s^2 + \vism^2}}\\
,\end{equation}
which gives us an indication for the timescale on which the accretion wake forms and the simulation can reach the theoretically expected Bondi-Hoyle accretion rate. This is roughly $10^3$ yr for $\vism = 1$ km/s, which is well within our simulation time of $10^4$ yr.

\section{Numerical set-up}\label{sec:setup_param}

\begin{table*}
\begin{center}
\caption{Initial conditions for our simulations. Only the density and velocity vary between simulations. The letter-number combinations behind the velocities and densities are used in the labels of the simulations.}
\label{tb:parameters_param}
\begin{tabular}{lrrrl}
\hline
\textbf{Parameter}&\textbf{Value}&&\textbf{Model}&\textbf{Description}\\
\hline
$\mathrm{N_{\rm disc}}$& 128 000&&&Number of disc particles\\
$\mu$&2.3&&&Mean molecular weight\\
$M_*$&$0.4\,\Msun$&&&\\
$M_{\rm disc}$&$0.004\,\Msun$&&&\\
$R_{\rm disc, inner}$&10\,AU&&&\\
$R_{\rm disc, outer}$&100\,AU&&&\\
$\Sigma(r)$&$\Sigma_0(\frac{r}{r_0})^{-1.5}$&&&Surface density profile\\
EoS&Isothermal&&&Equation of state\\
$T$&25\,K&&&Temperature of gas particles\\
$c_s$&$0.3\,\mathrm{km\, s^{-1}}$&&&Sound speed\\
$R_{\rm sink}$&1\,AU&&&Sink particle radius\\
$N_{\rm neighbours}$& $64\pm2$&&&\\
$\epsilon_{\rm grav}$&$1\,\mathrm{AU}$&&&Gravitational softening length\\
$\alpha_{\rm SPH}$&0.1&&&Artificial viscosity parameter\\
$L_{\rm cylinder}$& 400 AU&&&Length of computational domain\\
$R_{\rm cylinder}$& 200 AU&&&Radius of computational domain\\
$t_{\rm sim}$&10 000 yr&&&Timescale of the simulation\\
\hline
$\vism$&$1\, \mathrm{km\, s^{-1}}$&&V1&velocity of ISM\\
&$3\, \mathrm{km\, s^{-1}}$&&V3&\\
&$5\, \mathrm{km\, s^{-1}}$&&V5&\\
&$10\, \mathrm{km\, s^{-1}}$&&V10&\\
&$20\, \mathrm{km\, s^{-1}}$&&V20&\\
&$30\, \mathrm{km\, s^{-1}}$&&V30&\\
$n, \rho$&$5 \times 10^4 \, \mathrm{cm^{-3}}$&$1.9\times 10^{-19} \, \mathrm{g/cm^{-3}}$&N4&(Number) density of ISM\\
&$5 \times 10^5\, \mathrm{cm^{-3}}$&$1.9\times 10^{-18} \, \mathrm{g/cm^{-3}}$&N5&\\
&$5 \times 10^6\, \mathrm{cm^{-3}}$&$1.9\times 10^{-17} \, \mathrm{g/cm^{-3}}$&N6&\\

\hline
\end{tabular}
\medskip
\end{center}
\end{table*}

We test the theoretical model that we derived in Sect. \ref{sec:theory_param} by performing SPH simulations for different velocities and densities of the ISM.  We use the same set-up as in Paper I (see Fig. 1 therein), i.e. a disc perpendicular to an inflow of ISM. Although this perfect alignment may not be common in nature, it provides the best way to test our model. In Sect 5 we discuss that an inclination of the disc can also be incorporated in our model. Most of the parameters and initial conditions that we assume in this work, which are summarized in Table \ref{tb:parameters_param}, are the same as in Paper I. For a detailed discussion of these parameters, we refer to Sect. 3.1 of that paper. We discuss here which parameters we have changed and why. 

The simulations are set up and run within the AMUSE framework \citep{portegies_zwart13, pelupessy13}\footnote{\url{http://www.amusecode.org}}. We have chosen to use the SPH code Fi \citep{pelupessy04} instead of Gadget2 \citep{springel05,springel01} because the artificial viscosity parameter, $\alpha_{\rm SPH}$, is constant in Fi and therefore remains closer to the value that resembles the standard viscosity parameter $\alpha$ for protoplanetary discs\footnote{The performance issues we had with Fi in paper I turned out to be a problem with the most recent Fortran compiler.}. Fi uses the \citet{monaghan83} prescription for the viscosity factor $\Pi_{ij}$, but instead of the mean Fi takes the minimum of the density and maximum of the sound speed between two neighbours. Therefore Fi is able to resolve shocks even at a relatively low value of the artificial viscosity parameter $\alpha_{\rm SPH}$. As discussed in Sect.~\ref{sec:timescales_param}, numerical viscous effects become important and influence the outcome of our simulations after a few times $10^4$ year, depending on the density and velocity\footnote{To calculate the numerical viscous timescale, $H(r)$ in Eq. \ref{eq:disc-thickness_param} has to be replaced by $\langle h(r) \rangle$, the average smoothing length in the disc at that radius. The smoothing lengths in our simulations follow the curve of the scale height as a function of $r$ very well and are only larger by a factor of 2 at radii $\lesssim$ 10 AU. We therefore also use Eq. \ref{eq:tauvisc_param} to derive when accretion or viscous processes dominate the evolution of the disc numerically in our simulations.}. Therefore, and for computational reasons discussed in Sec. \ref{sec:velocities_param}, we let our simulations run to 10 000 years. 

To speed up the simulations, we increased the radius of the sink particle and the gravitational smoothing length, $\epsilon_{\rm grav}$, to 1 AU. Since the initial inner radius of the disc is 10 AU and we are mainly interested in the disc and star system as a whole, there is no need to resolve the star and disc on a scale smaller than 1 AU. We have checked that these changes do not affect the outcome of the simulations. We use the clump finding algorithm Hop \citep{eisenstein98} to distinguish between the disc and ISM flow in the parameterspace of $v_{\theta}$, $\rho$ and $v_x$; see Sect. 3.3 of paper I. We multiplied the values of all three quantities by a factor of 10 and squared the values of the low velocity runs ($\vism \le 10$ km/s) to sharpen the contrast in the parameter space, in particular at low velocities of the ISM. Once the disc is distinguished from the ISM, we determine its radius by calculating the column density of the disc and binning the values in concentric rings at each moment in time. The radius at which $\Sigma(r)\big|_{t} =\Sigma(100 {\rm AU})\big|_{t=0}$ is defined as the radius of the disc; see Sect. 3.3 and Fig. 3 of paper I for more details.

We start with a disc consisting of 128 000 SPH particles because this is the highest resolution at which we can perform a simulation in a reasonable amount of time. As we assume equal particle masses for both ISM and disc particles, this means that at the lowest number density of the ISM, $5 \times 10^4 \, \mathrm{cm^{-3}}$, the flow consists of roughly 500 particles when the computational domain is completely filled. At this resolution and for a simulation that runs up to $10^4$ years, an accretion rate of the order of $10^{-8}\,\Msun/$yr, which is the lowest rate we expect (see Fig. \ref{fig:accretion_param}), corresponds to the accretion of several thousand particles during a simulation. Therefore, even at this low resolution, the results we find are significantly above the threshold for noise due to low resolution.

The simulation is set up in such a way that the mass of an SPH particle is determined by quantities that vary per simulation, such as the density of the ISM, thickness of the slice of ISM added each time step, and the resolution. The latter is set by the number of disc particles, which is exactly equal for each simulation, and therefore the mass of the disc can vary slightly between simulations. We have chosen this set-up to make sure that the density of the ISM is exactly the same in all simulations to minimize the effect of small density fluctuations. We use the actual disc mass from the simulations to calculate the theoretical predictions.

\subsection{Initial conditions and parameters}

\subsubsection{Densities}\label{sec:densities_param}

We use three different number densities in our parameter study: $n = 5 \times 10^4,  5 \times 10^5$, and $5 \times 10^6\,\mathrm{cm}^{-3}$. The highest number density corresponds to the number density used in paper I, which is on the high end of the range of number densities expected in star-forming regions. Typically, the observed number density in star-forming regions is  $> 10^4\,\mathrm{cm}^{-3}$, while in the more quiet, i.e. non-star-forming parts of molecular clouds the number density is believed to be of the order of $10^2 - 10^3\,\mathrm{cm}^{-3}$ \citep[see e.g.][]{evans99, longmore13}. In our simulations we cannot decrease the number density below $10^4\,\mathrm{cm}^{-3}$ because the numerical resolution in our simulations would be too low as a result of the constraint of equal-mass SPH particles and the number of SPH particles that form the disc. Furthermore, at densities $< 10^4\,\mathrm{cm}^{-3}$, depending on the velocity, the disc may be in the regime where the accretion timescale is longer than the viscous timescale, as can be inferred from Fig. \ref{fig:timescales_param} for $\vism = 3$ k/s. We assume the mean molecular weight of the gas in our simulations is 2.3, as in paper I, yielding mass densities between $1.9 \times 10^{-19}$ and $1.9 \times 10^{-17}$ g/cm$^3$. The lowest density corresponds to the average density found in cores of embedded clusters; see e.g. Sect. 2.6 of \citet{lada03}. This assumes a homogeneous density distribution, while local inhomogeneities with higher densities can be expected. We show that even a brief transition through such a high density region has a large effect on the protoplanetary disc.

\subsubsection{Velocities}\label{sec:velocities_param}

The velocity dispersion of the Orion Nebula Cluster is measured to be 3.1 km/s \citep{furesz08}, meaning that individual stars can have higher velocities. Velocities are also  higher in more massive star-forming regions. Velocity dispersions in young globular clusters have been measured to range up to 30 km/s \citep{ostlin07, gieles10}. Apart from massive star-forming regions, relative velocities of the order of 10 km/s or more may also be expected from winds of supergiants and around ejecta from interacting binaries \citep[e.g.][]{kudritzki00,smith07}. The velocities we assume in this parameter study are 1, 3, 5, 10, 20, and 30 km/s, such that the velocity is roughly doubled at each step. Assuming a higher velocity leads to shorter internal time steps in the SPH code and thus the simulation becomes computationally more expensive. We have therefore not been able to let the simulations with 30 km/s run to 10 000 years. All simulations with 30 km/s have reached at least 5 000 years. 

For a velocity of 1 km/s, $R_{\rm BH}$ is larger than the radius of the computational domain. This means that we underestimate the theoretically expected Bondi-Hoyle accretion rate and the Bondi-Hoyle accretion wake cannot grow to its full length. Therefore, when we calculated the theoretically expected Bondi-Hoyle accretion rate for our simulations with $\vism = 1$ km/s, we used the radius of our computational domain, 200 AU, in Eq. \ref{eq:mdotBH_param} instead of the actual Bondi-Hoyle radius.

\section{Results}\label{sec:results_param}

For ease of reference, we label our simulations according to the variable parameters. For example, V1N6 corresponds to the simulation in which the ISM flow has a velocity of 1 km/s and a number density of $5 \times 10^6$ g/cm$^3$; see Table \ref{tb:parameters_param} for the abbreviations. We compare the various quantities as predicted by our model to those resulting from the simulations. While the number density is the quantity that we vary, our theoretical model requires a mass density to calculate the predicted evolution. We therefore refer to the mass density in the table and figures that contain theoretical values. We discuss in turn the evolution of the disc radius, accretion rate, change in disc mass, and surface density profile. We finish with the long-term predictions of our model in terms of the parameter space.

\subsection{Disc radius}\label{sec:res_radius_param}

\begin{figure*}[!ht]
\centering
\includegraphics[width=\textwidth]{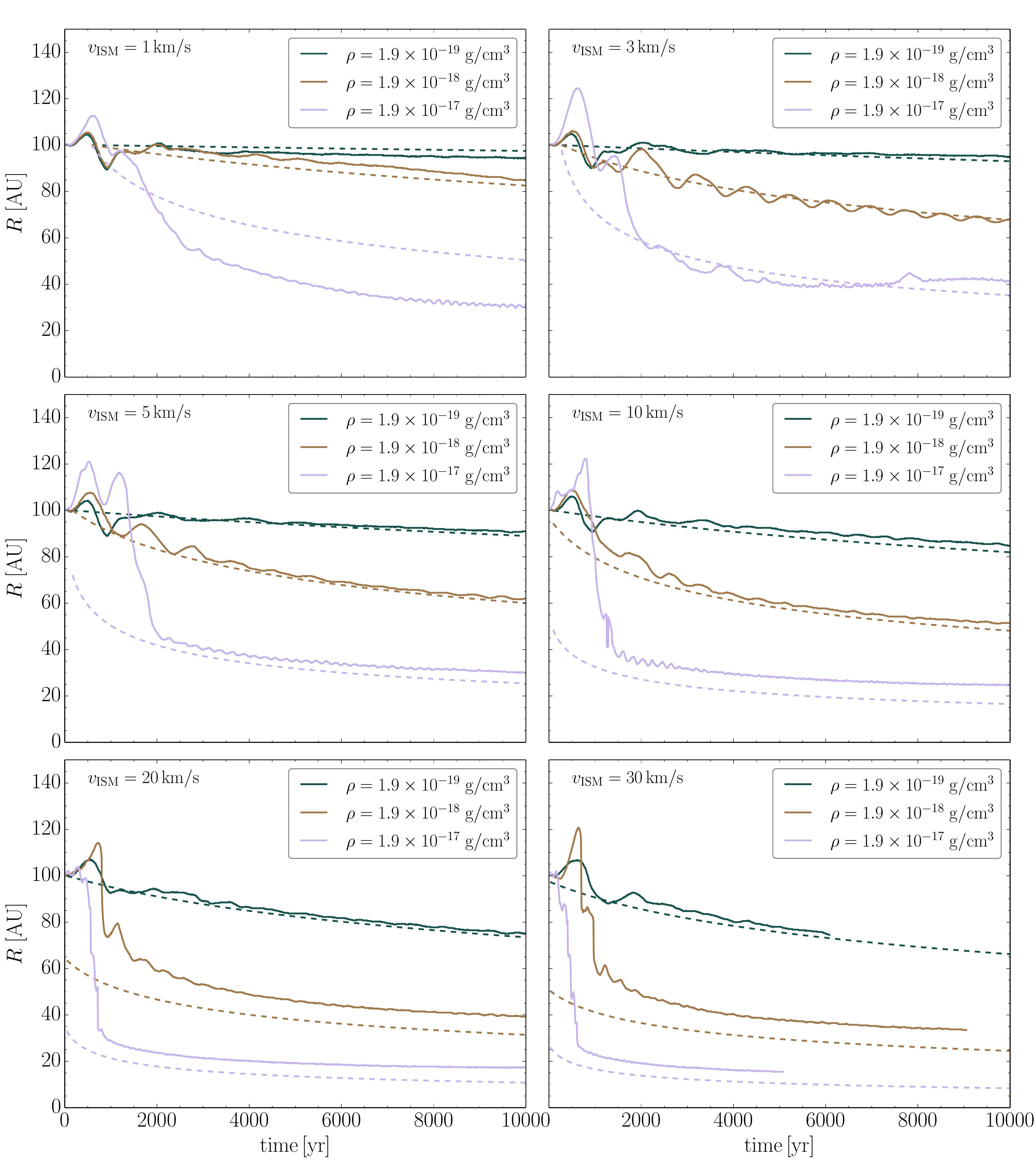}
    \caption{Disc radius (solid lines) as a function of time categorized by the velocity of the ISM in the simulation. The colour coding indicates the ISM density: green indicates $1.9 \times 10^{-19}$, brown $1.9 \times 10^{-18}$, and purple $1.9 \times 10^{-17}$ g/cm$^3$. The dashed lines in corresponding colours give our theoretical estimates of the disc radius (see Sect. \ref{sec:redistribution_param}). The simulations with $\vism = 30$ km/s were computationally expensive and have not reached 10 000 years.}\label{fig:radii_param}
\end{figure*}

Fig. \ref{fig:radii_param} shows the disc radius as a function of time for all our simulations together with our theoretical estimate. At the highest density, Hop has difficulty distinguishing the outer edge of the disc from the ISM at the start of the simulation for all velocities. For the densities in and around the edge of the disc it is hard to define a sharp boundary between the disc and ISM flow. When the outer parts of the disc have been stripped and/or contracted and a steady state is reached, Hop is able to properly discern the ISM flow from the disc. Some simulations show small fluctuations in the determination of the radius, notably simulation V3N5. During and after the stripping, the disc is not perfectly axially symmetric and the radius depends on how the surface density profile is averaged over the radii at the outer edge. These oscillations damp out in the long run and while they do, the equilibrium point follows the trend of the theoretical prediction. In simulations with a low density and velocity the disc has some time to spread viscously before the ISM flow reaches it and after first contact with the flow the disc contracts and then spreads towards an equilibrium state.

The effect of increasing the density by a factor of 10 is clearly visible at each velocity, in particular for $\vism \le 5$ because the starting radius is the same for all densities (i.e. stripping by ram pressure does not play a role). The increase in mass flux by a factor of 10 causes the disc to contract faster, as the specific angular momentum decreases faster. The effect is greatest in the beginning. The simulations seem to confirm what we find analytically: contraction occurs fastest when $r$ is large and slows down when $r$ has decreased for all densities and velocities (see Fig. \ref{fig:solution}). 

In general, our analytical model describes the evolution of disc radius with time from the simulations quite well. In some cases it somewhat underestimates the actual disc radius in the simulations, but this offset remains approximately constant in time and the rate at which the disc contracts is correctly reproduced by the analytical model. Simulation V1N6 is a notable exception in that the disc radius is overestimated by the model. In this case Bondi-Hoyle accretion becomes significant and the amount of accreted ISM is an order of magnitude larger than expected from purely face-on accretion, as discussed in Sect. \ref{sec:res_accretion_param}. Although most of the ISM is accreted directly onto the star through an accretion wake, ISM at radii larger than the disc radius are gravitationally focussed towards the disc and increase the theoretically expected mass flux. Thus the decrease in specific angular momentum is also more severe than when based on the analytical model.

Fig. \ref{fig:radii_param} also illustrates that our prediction for $R_{\rm trunc}$ is a reasonable estimate for the radius beyond which ram pressure dominates and disc material is stripped. In the simulations with $\vism \ge 10$ km/s at the highest density, the radius is reduced close to our estimated value around 1000 yr. The subsequent evolution of the disc radius in the simulation shows that we picked a proper starting point for the integration, although we consistently underestimate the radius of the disc. Eq. \ref{eq:Rtrunc_param} and the derivation by \citet{chevalier00} are both first order estimates, differing only by a constant factor of order unity. We could have changed this factor to match the final radii in our simulations, but this would have led to an overestimate of the accretion rate and the total disc mass, especially at lower densities (see Sect~\ref{sec:res_accretion_param} and \ref{sec:res_discmass_param}). Since the process of face-on accretion is most likely to occur at low densities, we decided not to alter this factor and to use Eq. \ref{eq:Rtrunc_param}, and consider our analytical derivation as a conservative estimate for the disc radii and accretion. 

In Fig. \ref{fig:Rcomp_param} we quantify how well our analytical model predicts the simulated disc radii by dividing the actual disc radius at the end of the simulation by our estimated radius. The colour scale shows the value of this ratio, which is also printed in the corresponding bin. We can distinguish three main regimes in Fig. \ref{fig:Rcomp_param}: (1) the regime where ram pressure truncates the disc radius to $< 100$ AU (upper right region above the $R_{\rm trunc} = 100$ AU line), (2) the regime where the Bondi-Hoyle radius is larger than the radius of the disc (below the horizontal $R_{\rm BH} = 100$ AU line), and (3) the region where the Bondi-Hoyle radius is smaller than the size of the disc and ram pressure does not decrease the initial disc radius below 100 AU (between both lines). Fig. \ref{fig:Rcomp_param} shows that in the regime where the ram pressure and Bondi-Hoyle accretion do not play a significant role, our analytical model predicts the simulated disc radius to within a few percent accuracy. Only at high density where ram pressure truncates the disc to a radius smaller than 100 AU or at low velocity where Bondi-Hoyle accretion becomes significant is our estimate off by at most 60\,\%. Since the estimate for the truncation radius is a first order approximation, it is not unexpected that the final radius shows a deviation from the actual disc radius in this regime. In general, we tend to underestimate the disc radius, except when the mass flux due to Bondi-Hoyle accretion is significant. In this regime, face-on accretion is insignificant compared to Bondi-Hoyle accretion as we discuss in the next section. 

It is interesting to point out that in some simulations, for example, V10N4 and V1N5, the mass flux ($\rhoism \vism$) is identical. The analytical estimate of the radius is therefore identical in both cases and we can see that it agrees to within 3\,\% with the disc radius in the simulations. 

\subsection{Entrained mass}\label{sec:res_accretion_param}

\begin{figure*}[!ht]
\centering
\includegraphics[width=\textwidth]{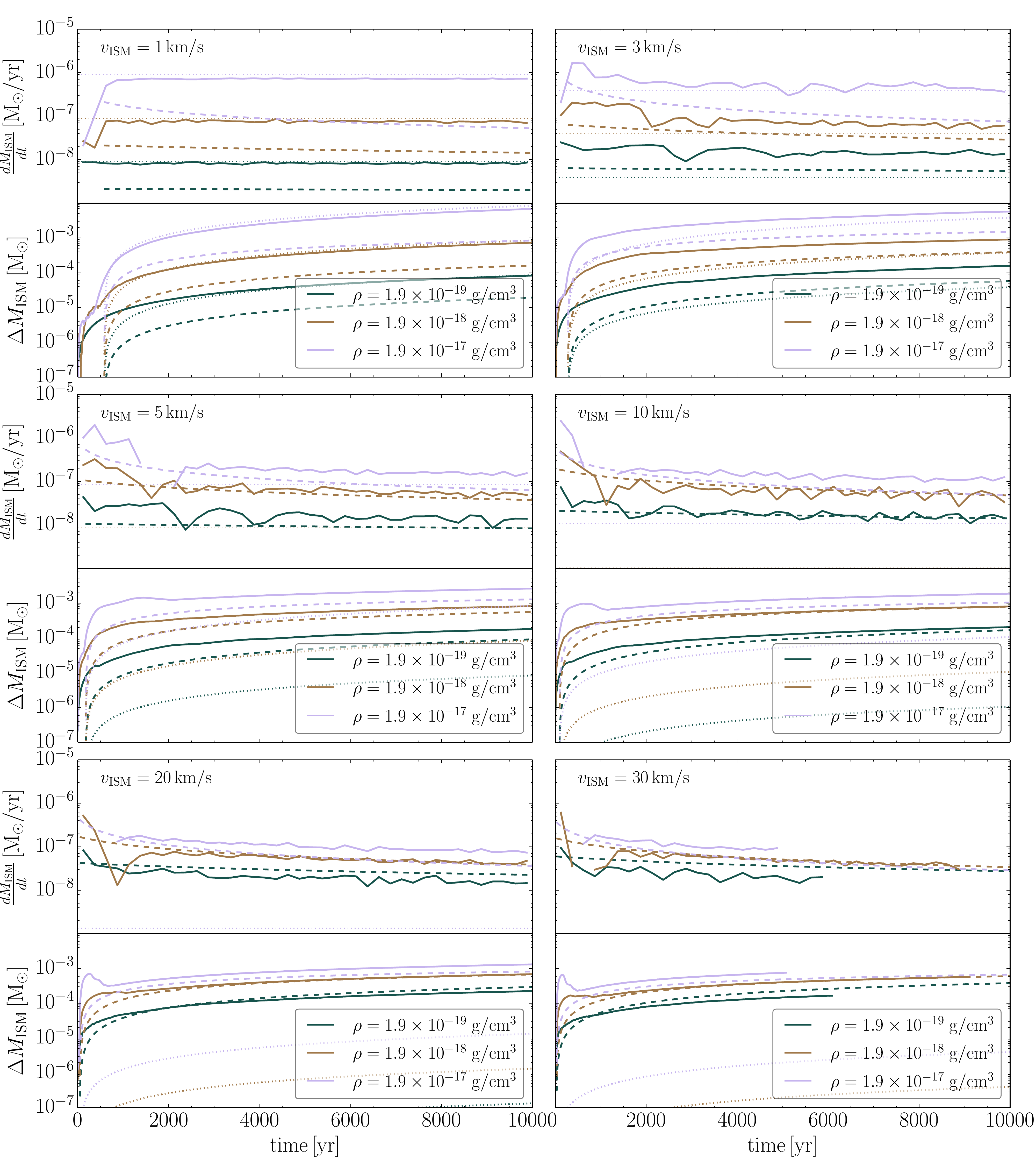}
    \caption{\emph{\bf{Upper half of each panel}} shows the accretion rates as a function of time found in our simulations (solid), expected from our theoretical model (dashed) and expected from Bondi-Hoyle accretion (dotted), respectively. The colour coding is as in Fig. \ref{fig:radii_param}. These rates are the derived from the quantity in \textbf{\bf{the lower half of the panel}}, which is the cumulative amount of ISM accreted by the disc and star (solid lines). Our theoretical estimate based on the size of the disc (see Eq. \ref{eq:dMdisk-dt}) is given by the dashed lines. The amount of accreted ISM expected from Bondi-Hoyle accretion is shown in dotted lines.}\label{fig:accretion_param}
\end{figure*}

Fig. \ref{fig:accretion_param} shows the amount of ISM accreted by the disc during the simulations in terms of both the cumulative accreted mass and the accretion rate. The lower part of each panel shows the amount of ISM, $\Delta M_{\rm ISM}$, that has been accreted by the disc and the star in solid lines. This value is calculated by adding the cumulative amount of ISM accreted onto the star and the amount of ISM that resides in the disc at each moment in time. We consider the star and the disc as one system by assuming that ISM that is entrained by the disc eventually ends up on the star. The dashed and dotted lines show the amount of accreted ISM expected from face-on and Bondi-Hoyle accretion, respectively. To calculate the Bondi-Hoyle accretion rate, we used the radius of our computational domain in Eq. \ref{eq:mdotBH_param}, which is 200 AU, if the theoretical Bondi-Hoyle radius exceeds this value. The top part of each panel shows the ISM accretion rate, i.e. the derivative of $\Delta M_{\rm ISM}$ averaged over intervals of 250 years. Likewise here solid, dashed, and dotted lines correspond to the rates from the simulation, face-on and Bondi-Hoyle accretion, respectively. 

To determine the amount of ISM that is associated with the disc, it is crucial that Hop is able to differentiate properly between the disc and ISM. As discussed in Sect. \ref{sec:res_radius_param}, Hop has some difficulty distinguishing the flow from disc when the flows hits the disc and the outer layers are stripped. These artefacts, which can be seen as bumps for the highest densities at $\vism \ge 5$ km/s in $\Delta M_{\rm ISM}$ in Fig. \ref{fig:accretion_param}, disappear when the simulation has reached a steady state. Unfortunately, it is not straightforward to discern the effects of Bondi-Hoyle accretion. For the simulations with $\vism=1$ and 3 km/s, in particular at the highest density, a Bondi-Hoyle accretion wake is formed, which is identified as disc material by Hop. Although the ISM in the accretion wake is not part of the disc, material in the accretion wake within the Bondi-Hoyle radius will eventually be accreted by the star, but necessarily by the disc. However, for $\vism = 3$ km/s the part of the accretion wake that is beyond the Bondi-Hoyle radius is still associated with the disc. Therefore the simulations suggest that more material has been accreted than expected from Bondi-Hoyle accretion. This is an unfortunate artefact of Hop but only creates an offset in the amount of accreted ISM. Hop consistently identifies the complete accretion wake with the disc and therefore the time derivative of $\Delta M_{\rm ISM}$ still provides a reliable measure for the rate at which the disc and star accrete ISM once a steady state is reached. Fig. \ref{fig:accretion_param} suggests that in most simulations the accretion of ISM starts immediately. This is also an artefact of Hop and therefore we choose a starting time for our analytical evolution model at the moment when the flow actually hits the disc. This creates another offset but as can be seen from the long-term trends, this initial offset is small compared to the final difference between our analytical model and the simulated result. 

It is clear, both from theory and from the simulations, that at $\vism = 1$ km/s the process of face-on accretion does not play a significant role in terms of ISM accretion onto the disc and star. The simulations reach the theoretically expected Bondi-Hoyle rate within a time that corresponds to the Bondi-Hoyle timescale we defined in Eq. \ref{eq:tau_BH_param}, which is roughly $10^3$ yr. During the time that the accretion wake is established, the accretion rate is lower than the theoretical rate. This partly explains why less material has been accreted than expected theoretically. Furthermore the accretion rate in our simulations is slightly lower than the theoretically expected rate. This may be caused by the fact that we used the radius of our computational domain instead of the theoretically expected Bondi-Hoyle radius and we are not correctly sampling the conditions at the boundary of the domain. In the simulations with $\vism = 3$ km/s, the Bondi-Hoyle radius equals 40\,\% of the radius of our computational domain and, in this case, i.e. model V3N6, the accretion rate is equal to, if not slightly higher than, the rate expected from Bondi-Hoyle accretion. At lower densities, i.e. models V3N4 and V3N5, we also find a rate that is higher than expected. The accretion rates appear to follow the theoretically suggested trend that first face-on accretion determines the rate at which ISM is accreted and after the disc has shrunk, Bondi-Hoyle accretion becomes the most efficient accretion process. This trend is also apparent in simulation V5N6. Although we describe the accretion as due to two different processes, they cannot really be separated in this regime; as the radius of the disc is of the order of the Bondi-Hoyle radius, the cross section with which the disc can sweep up ISM is effectively increased by gravitational focussing. If the disc radius becomes significantly smaller than the Bondi-Hoyle radius and the mass flux is sufficient, the gravitational focussing of ISM outside the disc radius may contribute to the contraction of the disc, as we discussed for $\vism = 1$ km/s in Sect. \ref{sec:res_radius_param}. This may also partly explain why the simulated accretion rate is higher than expected. As long as the disc radius is larger than the Bondi-Hoyle radius, i.e. in simulations with $\vism \gtrsim 5$ km/s, the accretion process is dominated by face-on accretion. In this regime, we tend to underestimate the accretion rate. However for all these cases we are also underestimating the radius of the disc and hence the surface area of the disc. Previous works, for example \citet{ouellette07} and paper I, suggest that the effective cross section of the disc is actually smaller than its physical size because the ISM flow is deflected around the outer edge of the disc. Apparently we also underestimate the \emph{effective} radius of the disc except in the simulations with $\vism = 20$ and 30 km/s at the lowest density where our estimate of the disc radius approximates the actual radius very well. In these cases, we overestimate the effective cross section of the disc and hence the accretion rate. On the other hand, at $\vism = 5$ and 10 km/s our estimate also agrees very well with the actual disc radius and in these cases we are not underestimating the accretion rates. This may be related to the ram pressure, which has a stronger influence on the outer edge of the disc at high velocities, $\vism \ge 20$ km/s. We find that the disc loses some of its initial material due to stripping (see Sect. \ref{sec:res_discmass_param}) and this would make it less efficient at sweeping up ISM in its outer regions. However, the underestimate is most likely a result of the low resolution of ISM particles in the SPH code, which is more pronounced when the radius of the disc is smaller. We address this issue in Sect. \ref{sec:dis_resolution_param}.

\begin{figure*}[!ht]
        \centering
        \subfloat[(a)]{\includegraphics[width=0.5\textwidth]{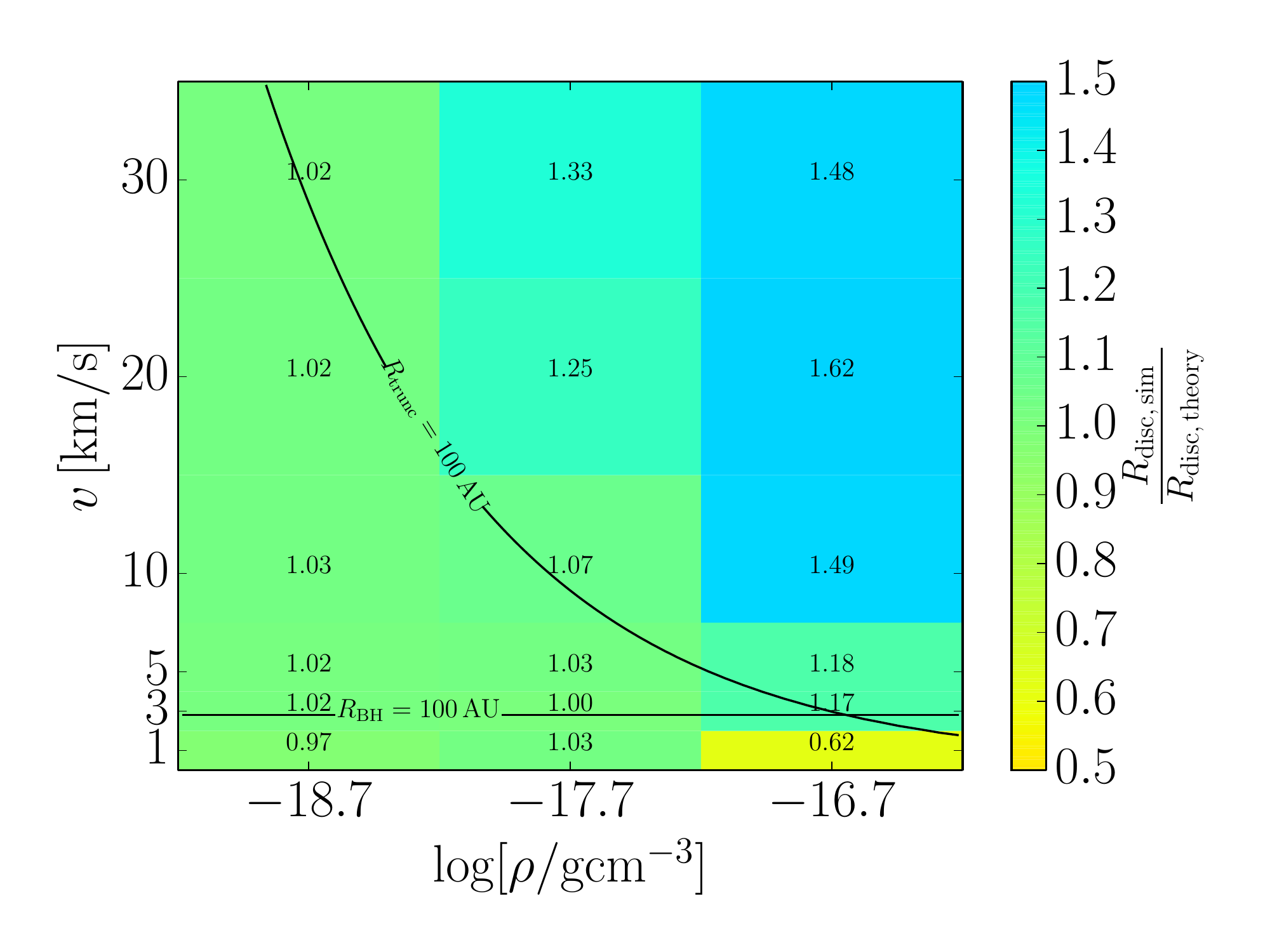}\label{fig:Rcomp_param}}
        \subfloat[(b)]{\includegraphics[width=0.5\textwidth]{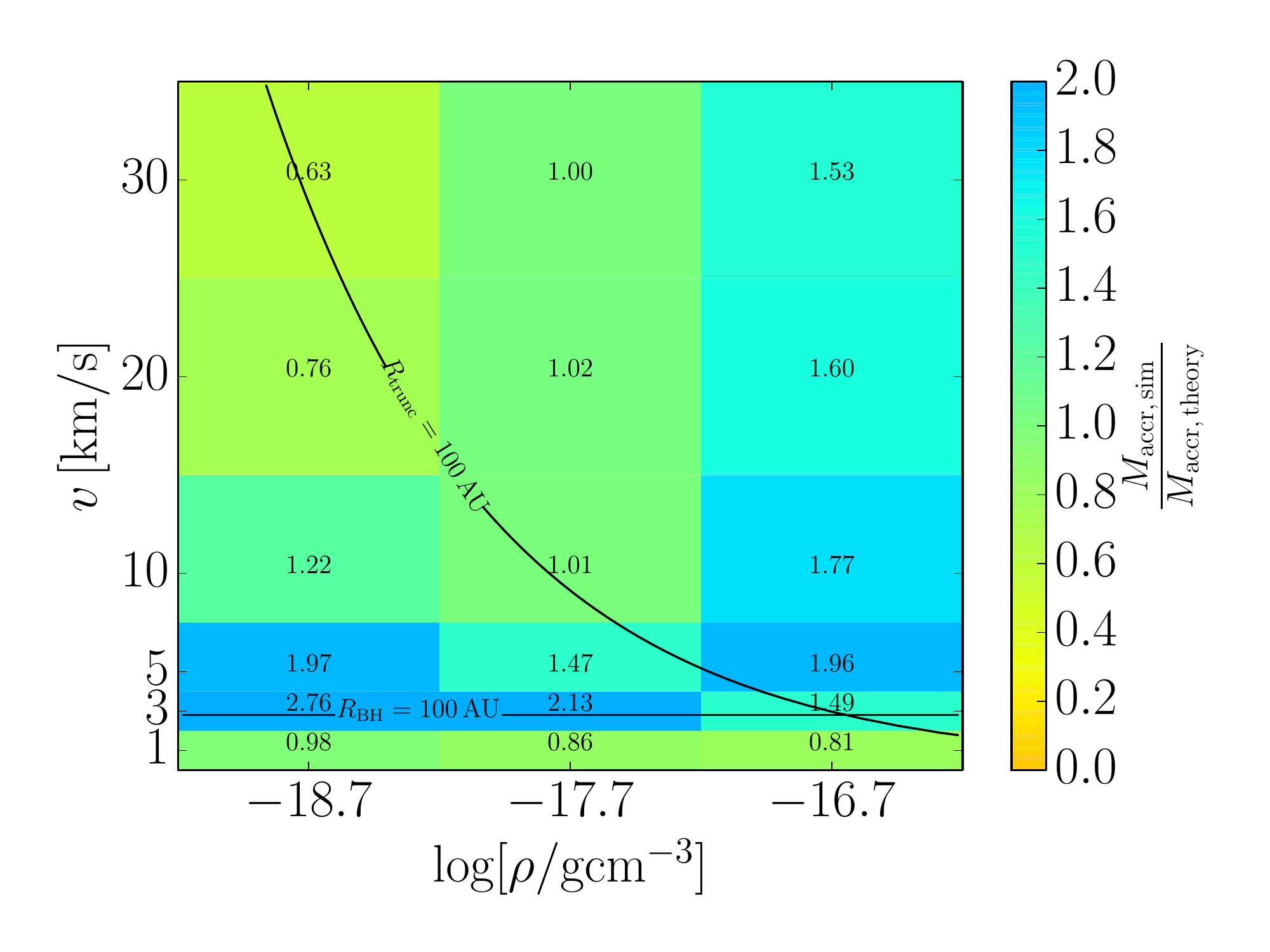}\label{fig:Mcomp_param}}\\
        \subfloat[(b)]{\includegraphics[width=0.5\textwidth]{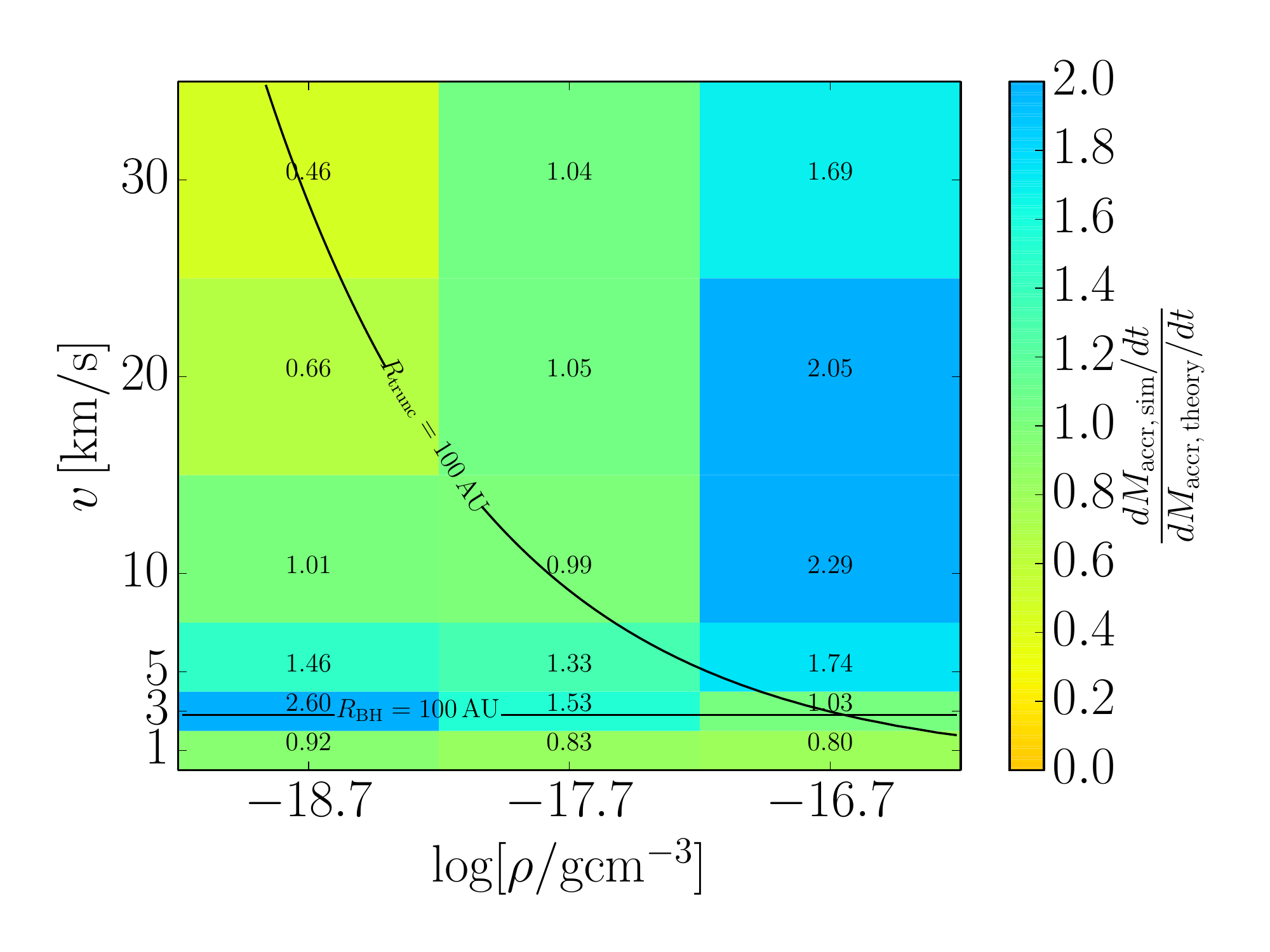}\label{fig:Mdotcomp_param}}\begin{minipage}[b]{0.5\textwidth}
    \caption{Measures of the performance of our theoretical model. \textbf{\emph{(a):}} The radius of the disc at the end of the simulation divided by the predicted radius at that time. \textbf{\emph{(b):}} The amount of ISM accreted by the disc and star at the end of the simulation divided by the theoretically expected value at that time. To calculate the expected amount, we added the additional contribution of Bondi-Hoyle accretion from the moment the Bondi-Hoyle rate is higher than the geometric rate. \textbf{\emph{(c):}} The accretion rate at the end of the simulation divided by the theoretically expected rate at that time. Here we  took the maximum of the geometric and Bondi-Hoyle accretion rate at each moment and averaged both the theoretical and simulated values over the last 1 000 years of the simulation. The lines indicate the regime where the Bondi-Hoyle radius and truncation radius are larger (below) and smaller (above), respectively, than the initial radius of the disc (see Sect. \ref{sec:res_radius_param}). The values in the grid cells correspond to the colour coding.}\end{minipage}\\%
\end{figure*}

We summarize the performance of our analytical model in predicting the accretion rates and accreted mass in Fig. \ref{fig:Mcomp_param} and \ref{fig:Mdotcomp_param}, where we show the ratio of the total accreted ISM and accretion rates from the simulations to their predicted values. To calculate the predicted value of the total amount of accreted ISM, we used Eq. \ref{eq:deltaM_param} and added the difference between the Bondi-Hoyle accretion rate and geometric rate (Eq. \ref{eq:dMdisk-dt}) integrated over time from the moment the former is larger than the latter. The theoretical accretion rate in Fig. \ref{fig:Mdotcomp_param} is determined by taking the maximum of the face-on and Bondi-Hoyle accretion rate at each moment. We averaged both the theoretical and simulated accretion rate over the last 1 000 years of the simulations in order for the result not to be dominated by short timescale fluctuations. 

In the regime in which part of the Bondi-Hoyle accretion wake is erroneously considered part of the disc by Hop, i.e. $\vism = 3$ and 5 km/s, the accretion rate in Fig. \ref{fig:Mdotcomp_param} provides a better indication for the performance of our model. The agreement is not as good as for the radius (Fig. \ref{fig:Rcomp_param}), but is still roughly within a factor of two. We overestimate the accretion rate in two cases, i.e. V20N4 and V30N4. As pointed out above, this is probably a numerical artefact and is further discussed in Sect. \ref{sec:dis_resolution_param}. The disagreement is worst for simulations were we underestimate the accretion rate. Therefore, considering the numerical effects in simulation V20N4 and V30N4, Figs. \ref{fig:Mcomp_param} and \ref{fig:Mdotcomp_param} indicate that our parametrization can be used as a conservative estimate for the amount of ISM accreted by a protoplanetary disc moving through ambient gas.

\subsection{Change of the disc mass}\label{sec:res_discmass_param}

\begin{figure*}[!ht]
\centering
\includegraphics[width=\textwidth]{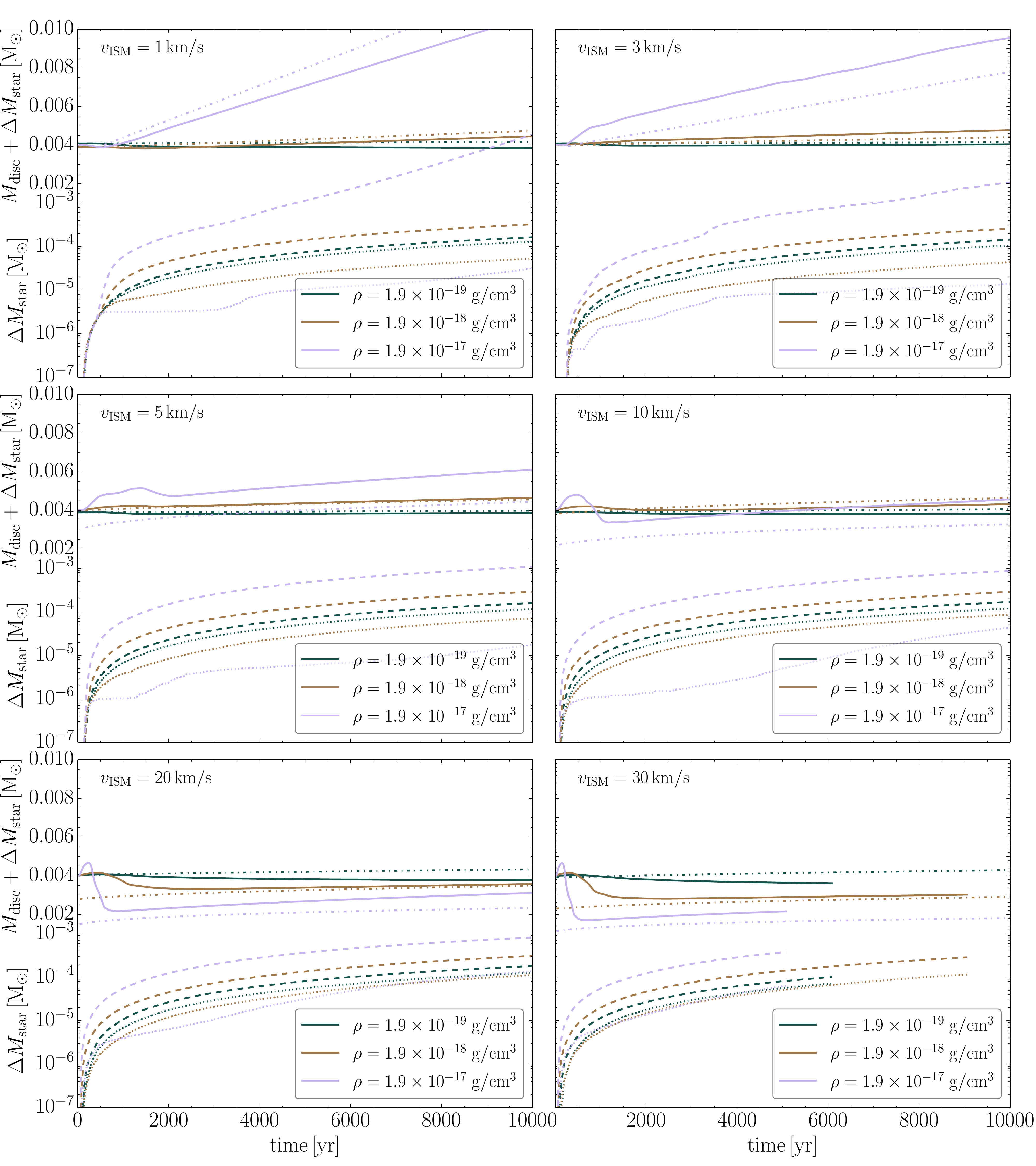}
 \caption{\emph{\bf{Top part of each panel}} shows the mass of the disc to which we added the total amount of material accreted onto the star as a function of time (solid lines) categorized by the velocity of the ISM in the simulation. The colour coding is as in Fig. \ref{fig:radii_param}. Our theoretical estimate of the same quantity (see Sect. \ref{sec:res_discmass_param}) is given by the dash-dotted lines in corresponding colours. \emph{\bf{The bottom part of each panel}} shows the total amount of material accreted onto the star in dashed lines with the corresponding colour. The dotted lines give the amount of initial disc material that is accreted onto the star. The y-scale is continuous because all quantities have the same unit.}\label{fig:discmass_param}
\end{figure*}

The panels in Fig. \ref{fig:discmass_param} show two quantities on the same y-axis. The top parts of each panel shows the sum of the current mass of the disc and total mass (both ISM and initial disc material) that has been accreted onto the star as solid lines. We refer to this quantity as the system mass. The dash-dotted lines show our theoretical estimate for the system mass. We treat the disc and star as a whole and compare to this quantity because our analytical model does not take into account mass lost from the disc due to accretion onto the star. To calculate the system mass, we used Eq. \ref{eq:Mdisc_param} and to this we added the mass gained from the difference between the Bondi-Hoyle accretion rate and the face-on accretion, only during time intervals when the theoretical Bondi-Hoyle accretion rate was higher. Fig. \ref{fig:accretion_param} shows that this difference is only relevant for velocities $\le 3$ km/s. As discussed in Sect. \ref{sec:setup_param}, the initial mass of the disc can vary slightly between simulations with different densities. This does not affect our results because for the theoretical calculations we used the initial disc mass in the simulation. The bottom part of each panel in Fig. \ref{fig:discmass_param} shows the cumulative amount of material that has been accreted onto the star, both ISM and initial disc material, as dashed lines. The dotted lines give the amount of initial disc material that has been accreted onto the star. The difference between the dashed and dotted lines thus gives the amount of ISM accreted onto the star.

We first discuss the simulations with the highest density. For $\vism = 1$ and 3 km/s Bondi-Hoyle accretion dominates the accretion process, as discussed before. Considering the shortcomings of Hop in this regime, it is not unexpected that our theoretical estimate for the system mass is off. The offset can mostly be explained, however, by the difference between the actual and theoretical accretion rates and the fact that the accretion wake is considered part of the disc. For $\vism \ge 5$ km/s Hop at first overestimates the system mass. Once ISM that has collided with the outer parts of the disc is dragged along with the flow and is no longer considered to be associated with the disc, a steady state is reached. For these velocities, $\vism \ge 5$, we also see an offset between the predicted and simulated system mass but this is caused by the amount of predicted stripped material. This offset is not caused by Hop because we checked the disc mass by drawing a fictitious cylinder around the disc and summing all the mass within that cylinder, which gives the same result and long-term trend. The reason for the discrepancy is twofold: (1) our derivation for the truncation radius is only a first order estimate and (2) in the simulations the cut-off in the surface density profile is not as sharp as we assume in our theoretical model. Both effects combined lead to an overestimate of the amount of stripped disc material. This overestimate may be less severe when the truncation radius is smaller. The initial surface density profile is steeper at smaller radii and our estimate for the largest annulus of the disc that is able to resist ram pressure may therefore deviate less from the actual value than for larger truncation radii, where the surface density profile is less steep. As discussed before, we could chose the truncation radius such that the stripped mass found in the simulations agrees with our model. Unfortunately, there is no constant factor we could use in Eq. \ref{eq:Rtrunc_param} that simultaneously improves the estimate of all quantities in Figs. \ref{fig:radii_param}, \ref{fig:accretion_param}, and \ref{fig:discmass_param}. Furthermore, an increase of the truncation radius would lead to an overestimate of the accretion rate onto the disc and we would no longer be able to use our model as a conservative estimate on longer timescales, at least until viscous effects can no longer be neglected.

We find the best agreement between our model and the simulations for $\rhoism = 1.9 \times 10^{-18}\,\mathrm{g/cm^3}$. In this regime ram pressure stripping only plays a significant role for $\vism \ge 20$ km/s and even in these cases there is fairly good agreement between our model and simulations. Our estimate for the truncation radius for these
cases is relatively small and might be more accurate, as discussed above. For V1N5 and V10N5 we overestimate the system mass and this is even more severe in all simulations at the lowest density. In fact, we see that the system as a whole actually loses mass in all simulations at the lowest density. This can only mean that material from the disc is being stripped continuously.

\begin{table*}
\begin{minipage}{\textwidth}  
\centering
\caption{Percentage of initial disc mass that is stripped from the disc over the course of the simulation. The first number in each column is the percentage found in the simulation, the second number is the theoretical prediction. Columns denote the velocity of the simulation and rows the density.}
{\begin{tabular}{l|cccccc}
$\rhoism$\hfill\raisebox{0.5 em}{$\vism$}&$1$\,km/s&$3$\,km/s&$5$\,km/s&$10$\,km/s&$20$\,km/s&$30$\,km/s\\
\hline
$1.9 \times 10^{-19}\,\mathrm{g/cm^3}$&$8.0\big/ 0$\%&$5.0\big/0$\%&$5.2\big/0$\%&$6.9\big/0$\%&$12.1\big/0$\%&$13.7\big/2.0$\%\\
$1.9 \times 10^{-18}\,\mathrm{g/cm^3}$&$4.4\big/0$\%&$2.9\big/0$\%&$4.5\big/0$\%&$12.3\big/4.1$\%&$27.8\big/29.7$\%&$38.8\big/42.4$\%\\
$1.9 \times 10^{-17}\,\mathrm{g/cm^3}$&$3.7\big/0$\%&$4.4\big/1.5$\%&$13.1\big/22.0$\%&$32.9\big/44.4$\%&$55.1\big/62.2$\%&$65.0\big/71.6$\%\\
\end{tabular}
}
\label{tb:strippeddisc_param}
\end{minipage}
\end{table*}

In Table \ref{tb:strippeddisc_param} we summarize the fraction of initial disc material that is lost from the disc during the simulation but not accreted onto the star. The left value in each column is the percentage of stripped material from the simulation and the right number is the theoretical value. The latter is calculated from the initial mass of the disc outside the theoretical truncation radius. At the highest densities and velocities we overestimate the amount of stripped material, as discussed above. However, at the lowest density and at low velocities the disc loses material; this is not accounted for in our model. 
At $\vism \le 3$ km/s, there appears to be a trend with decreasing density at a given velocity: the lower the density, the higher the amount of mass loss that is not accounted for by our model. Fig. \ref{fig:discmass_param} shows that more initial disc material is accreted onto the star in the simulation with the lowest density. The more initial disc material is accreted onto the star and lost from the disc, the more the disc has to spread outwards to preserve its angular momentum. As can be seen in Fig. \ref{fig:surfdens_param} for simulation V1N4, the disc extends further outwards and the edge is less well defined than for simulations with a higher density and velocity. In this case, 9\,\% of the disc mass is actually outside the disc radius. This implies that viscous spreading of the disc is more severe at lower densities, making it more vulnerable to ram pressure stripping. In particular, at a lower density the mass flux onto the disc is smaller and hence the disc contracts less, which otherwise counteracts the effect of viscous spreading. However, it could also simply be an artefact of the low resolution of the ISM particles. We address this in Sect. \ref{sec:dis_stripping_param}.

At the lowest density, the unaccounted stripping appears to increase with velocity, not taking V1N4 into account. This trend with velocity suggests that ram pressure is the culprit of the mass loss. In paper I we also found that the disc was continuously losing mass from the outer edge. In that work, our results suggest that the stripping of the disc might to a large extent be a numerical effect; see Fig. 6 in paper I, where for a simulation with $\vism = 20$ km/s and $\rhoism = 1.9 \times 10^{-17}\,\mathrm{g/cm^3}$ the amount of stripped material decreases with increasing resolution. Furthermore, other work, for example \citet{moeckel09}, found no significant mass loss from the disc. We have seen in Sect. \ref{sec:res_radius_param} and \ref{sec:res_accretion_param} that our model 
provides a conservative estimate for the disc radius and accretion rate. Therefore we argue that this mass loss rate does not affect the predictive power of our model. We discuss this issue further in Sect. \ref{sec:dis_stripping_param}.

\subsection{Evolution of the surface density profile}\label{sec:res_surfdens_param}

\begin{figure*}[!ht]
        \centering
        \subfloat[(a)]{\includegraphics[width=0.5\textwidth]{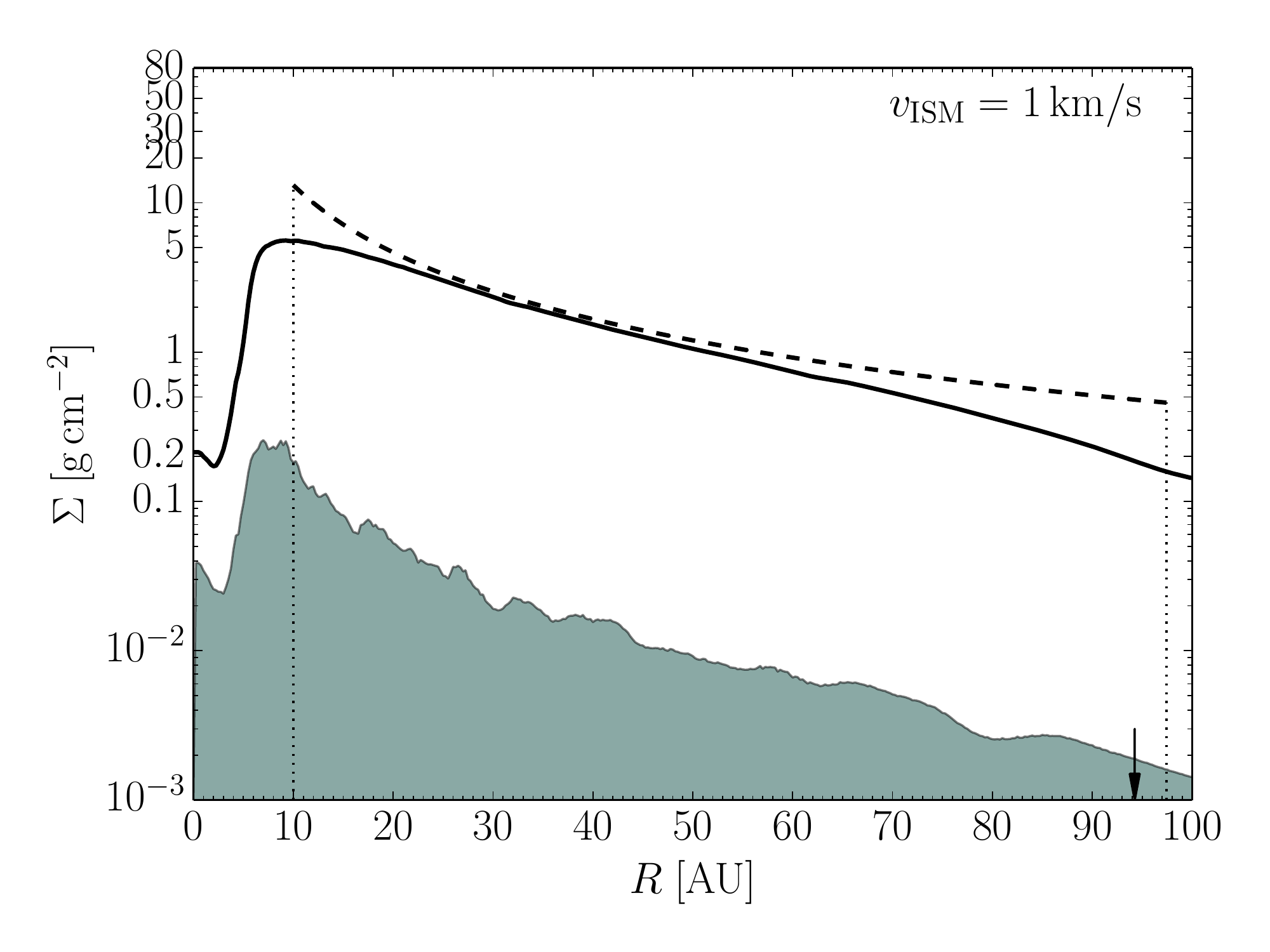}\label{fig:surfdensv1n5e4_param}}
        \subfloat[(b)]{\includegraphics[width=0.5\textwidth]{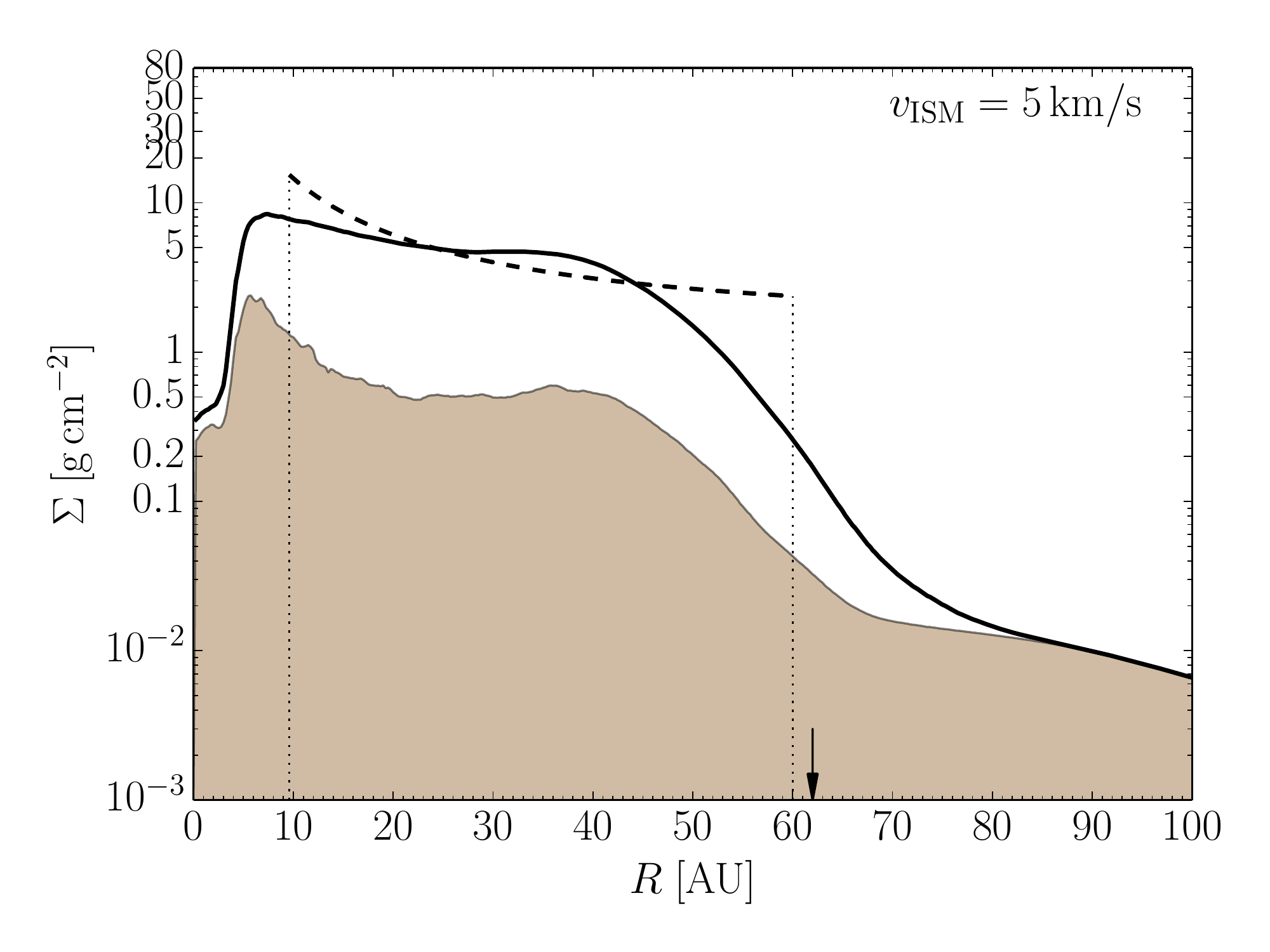}\label{fig:surfdensv5n5e5_param}}\\
        \subfloat[(b)]{\includegraphics[width=0.5\textwidth]{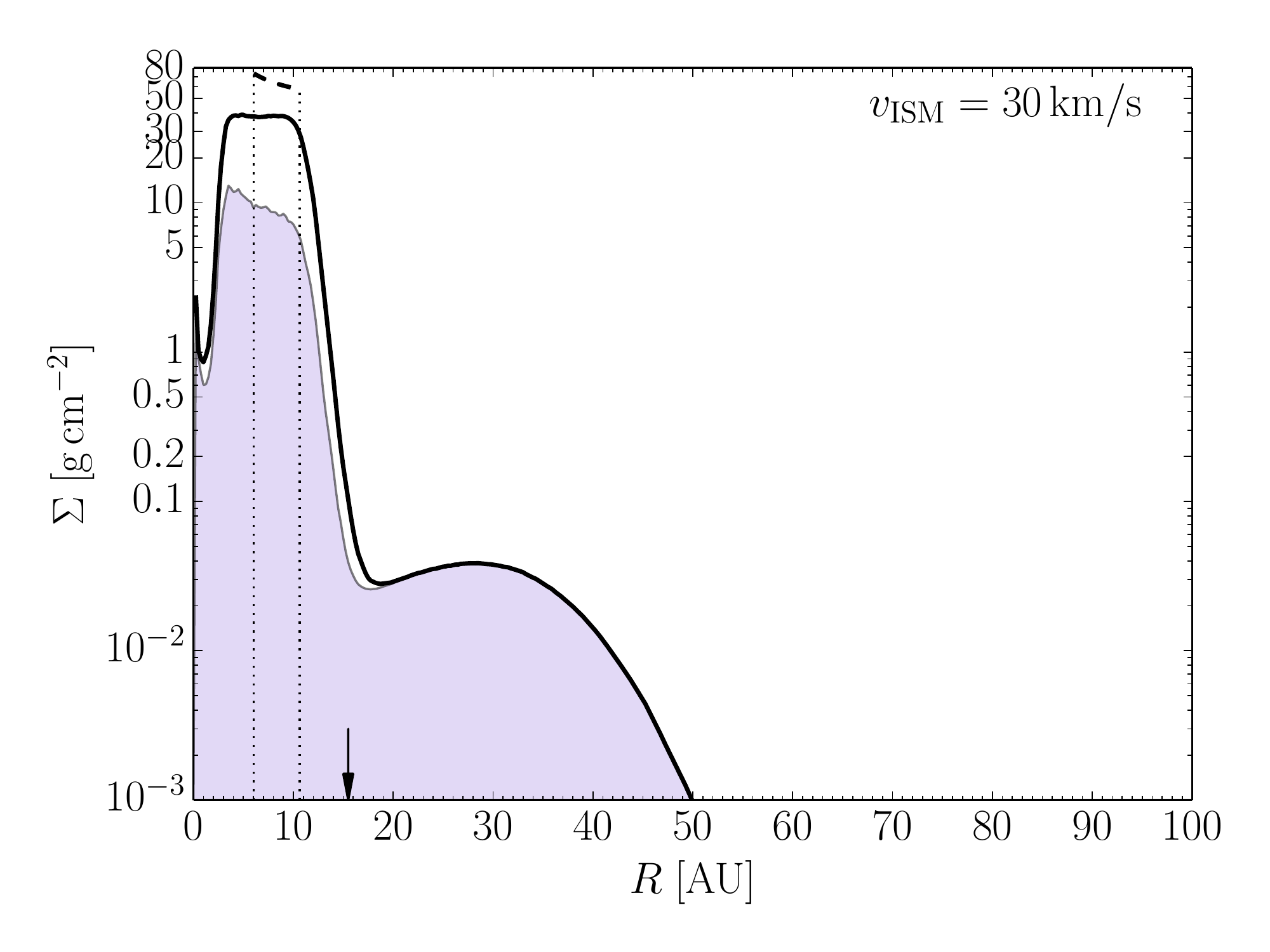}\label{fig:surfdensv30n5e6_param}}\begin{minipage}[b]{0.5\textwidth}
    \caption{Surface density profile at the end of each simulation (solid line) and as predicted by our model (dashed line) for simulations \textbf{\emph{(a)}} V1N4, \textbf{\emph{(b)}}  V5N5, and \textbf{\emph{(c)}} V30N6. The dotted vertical lines indicate the inner and outer radius of the disc according to our model and the arrow indicates the radius of the disc as defined in the simulation. The shaded area illustrates the contribution of the ISM to the surface density profile, where the colour corresponds to the density: green indicates $1.9 \times 10^{-19}$, brown $1.9 \times 10^{-18}$, and purple $1.9 \times 10^{-17}$ g/cm$^3$ .}\label{fig:surfdens_param}\vspace{3cm}\end{minipage}\\%
\end{figure*}

Fig. \ref{fig:surfdens_param} shows the surface density profiles of three simulations: V1N4, V5N5, and V30N6. We have chosen these simulations as a representative sample of the parameter space. The main features expected from our model are visible: as the mass flux increases, the disc shrinks and the surface density profile both flattens and increases. The flattening is seen in all simulations with $\vism \ge 3$ km/s and $\rhoism \ge 1.9 \times 10^{-18}$ g/cm$^3$. As may be expected, the disagreement is strongest at the inner and outer edge of the disc: inward and outward viscous spreading, respectively, decrease the surface density profile at each end. The ISM that has been swept up by the disc is not distributed evenly across the disc but accumulates preferentially at small radii. This is most clearly visible in Fig. \ref{fig:surfdensv1n5e4_param} for simulation V1N4. The ISM originally entered the computational domain at larger radii and migrated inwards through the disc. This accumulation of ISM at small radii occurs in all simulations to some extent. In our model, we assumed that when the ISM is entrained by the disc it co-rotates instantaneously with the disc material at that radius. However, as a result of the finite viscosity between the swept-up ISM and rotating disc material, the ISM moves to a tighter orbit before it achieves co-rotation. This partly explains why (1) the contribution of ISM to the surface density profile is not evenly distributed across the disc and (2) the contribution of ISM drops more sharply at the edge of the disc. 

Substantial flattening of the surface density profile is seen in the two simulations with a higher density (Figs. \ref{fig:surfdensv5n5e5_param} and \ref{fig:surfdensv30n5e6_param}). The flattening is more prominent at the highest mass flux (simulation V30N6) and is more pronounced than predicted by our model. The second bump in the surface density profile around 30 AU, seen in Fig. \ref{fig:surfdensv30n5e6_param}, can be attributed to ISM, which is considered part of the disc by Hop while it actually flows around the outer edge. This bump only accounts for 1\,\% of the disc mass and does not affect the conclusions reached in Sect. \ref{sec:res_discmass_param}. In the V30N6 simulation the surface density profile is approximately a factor 2 lower than expected. However, since the surface density profile in the simulation extends to a larger radius than expected, the total mass in the disc is still somewhat larger than predicted. 

We neglected viscous effects in our model and it is obvious from Fig. \ref{fig:surfdens_param} that these effects are present, both numerically and probably also physically. The latter most likely plays a role at the inner and outer edge of the disc in our simulations. Although the surface density profile shows clear quantitative differences compared to the predicted behaviour, the profile still qualitatively evolves as expected. Furthermore, the global predictions of the model, i.e. the radius of the disc and accretion rate, also quantitatively follow the predicted evolution. In general, the agreement is satisfactory, especially considering the approximate nature of the theoretical model, on one hand, and the limitations of the numerics, on the other.

\subsection{Long-term effects of ISM accretion}\label{sec:res_predictions_param}

\begin{figure*}[!ht]
        \centering
        \subfloat[(a)]{\includegraphics[width=0.5\textwidth]{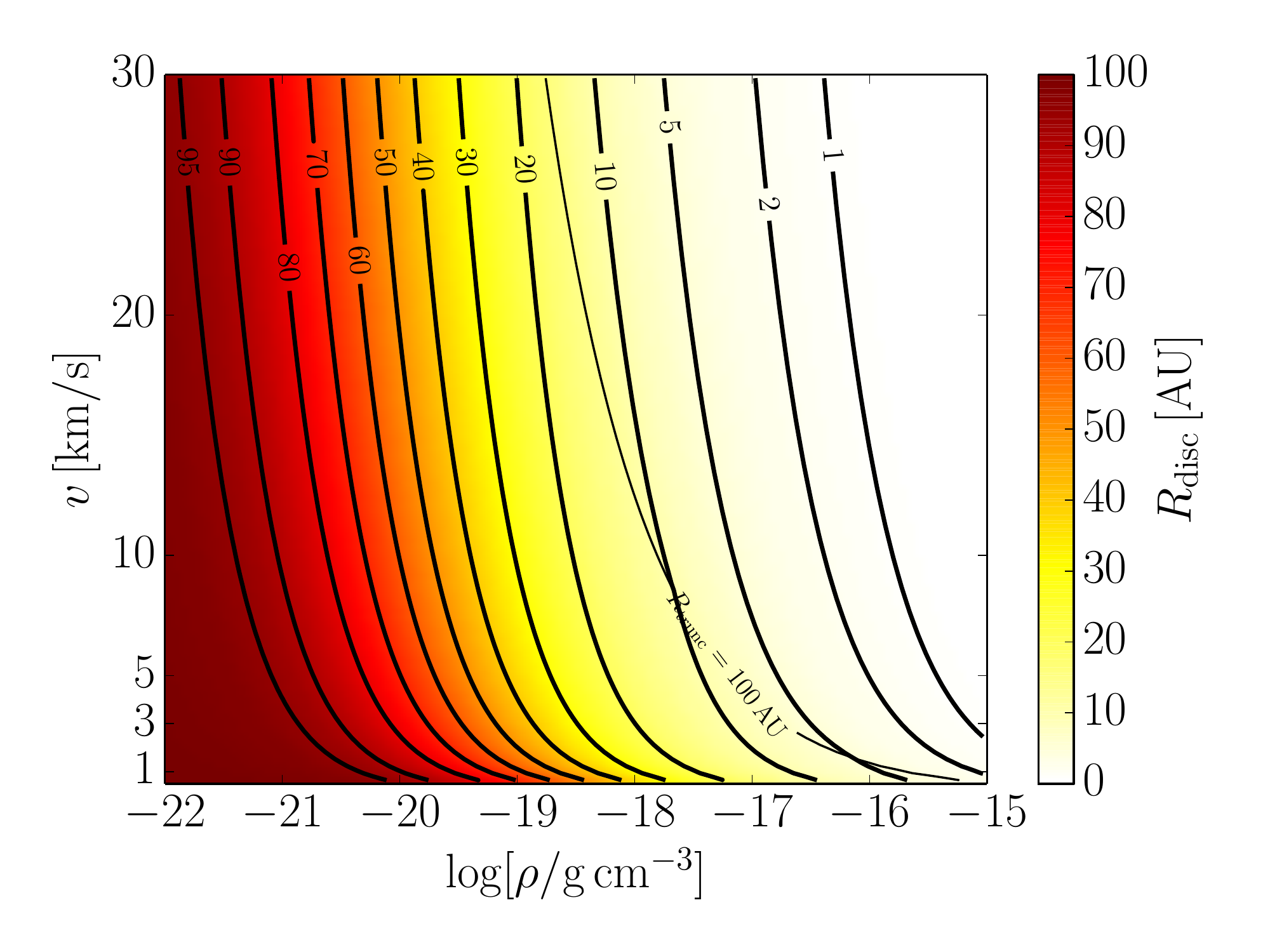}\label{fig:Rend_param}}
        \subfloat[(b)]{\includegraphics[width=0.5\textwidth]{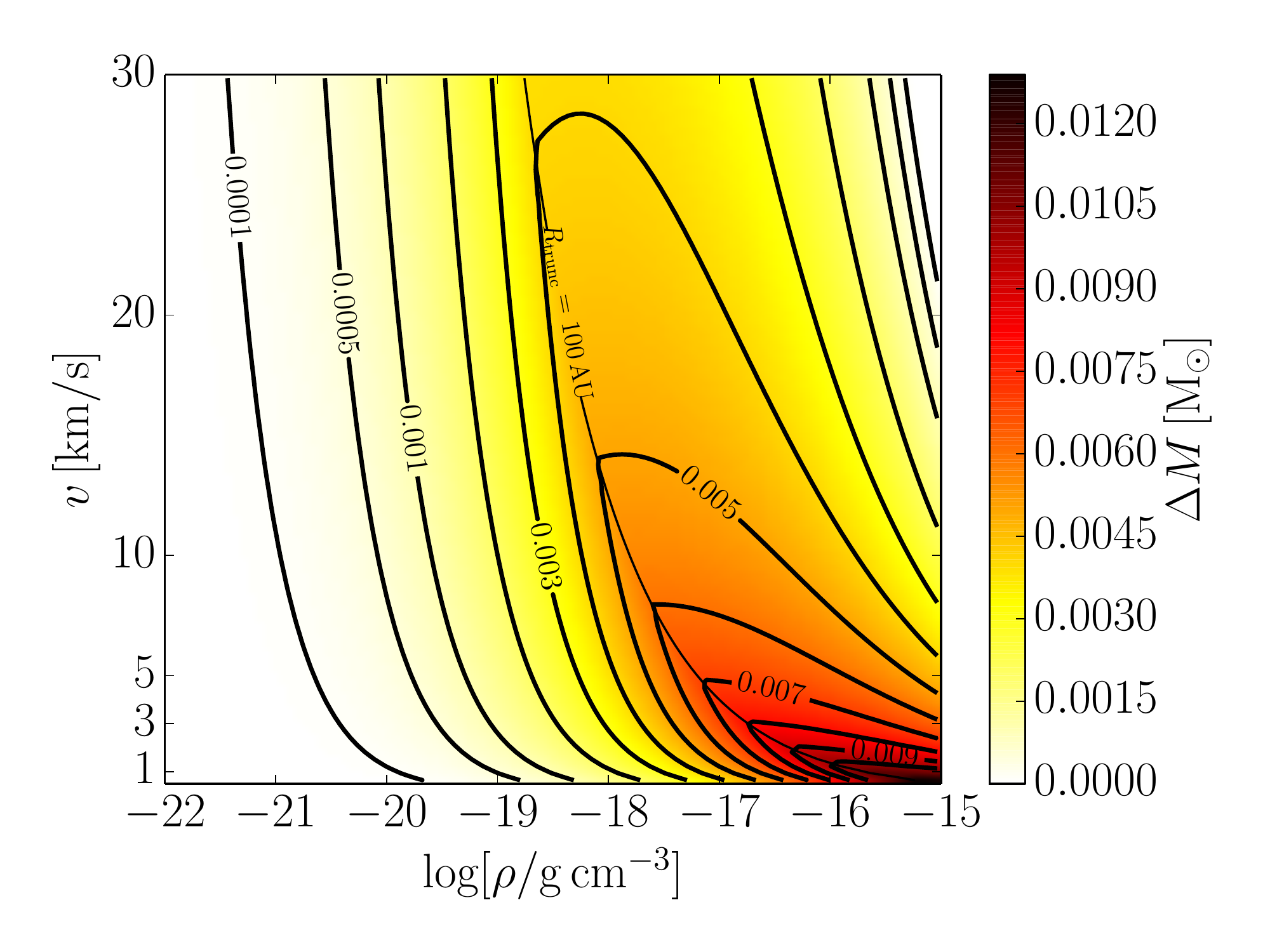}\label{fig:Mend_param}}
    \caption{Predictions of our theoretical model as a function of the density-velocity parameter space for \textbf{\emph{(a):}} the radius of the disc and \textbf{\emph{(b):}} the amount of accreted ISM after continuous accretion for 1 Myr. Bondi-Hoyle accretion is not taken into account.}  
\end{figure*}

In the hypothetical case in which the discs in our simulations manage to accrete continuously on a timescale much longer than $10^4$ yrs, we can use our theoretical model of Sect. \ref{sec:model_param} to calculate the implications for the sizes of the discs and amount of accreted ISM. Of course, protoplanetary discs will never  be embedded continuously for such a long time in a flow of constant density and velocity. However, our model only depends on the time-integrated mass flux through the parameter $\tau$ (Eq. \ref{eq:tau-integral}). It is questionable to neglect the viscous evolution of the disc on such a long timescale, but the model gives an indication of the effects on the long term. 

The embedded phase in the evolution of young clusters is believed to last of the order of 1-3 Myr \citep{lada03, portegies_zwart10}. Figs. \ref{fig:Rend_param} and \ref{fig:Mend_param} therefore show the disc radius and amount of accreted ISM after 1 Myr of continuous accretion as a function of the ISM density and velocity. We did not account for Bondi-Hoyle accretion in these figures. As long as ram-pressure stripping is not important (i.e. to the left of the line $R_{\rm trunc} =  100$ AU) the contours of constant radii are lines of constant $\rhoism \vism$ because our theoretical model scales with $\tau \propto \rhoism \vism$; see Eq. \ref{eq:tau-integral}. We have shown in Sect. \ref{sec:model_param} that in the long term the disc radius scales with $\tau^{-2/5}$ and the disc mass and, thus, $\Delta M$ scale with $r_0^2 \tau^{1/5}$, where $r_0$ is a typical initial radius inside the disc. Fig. \ref{fig:Rend_param} confirms that the radius decreases with increasing density and velocity. At low densities and velocities, where ram pressure does not truncate the disc and thus $r_0$ is constant, $\Delta M$ increases with increasing density and velocity. However, when ram pressure does truncate the disc, then $r_0^2 \propto R_{\rm trunc}^2 \propto \rhoism^{-4/7} \vism^{-8/7}$ for an initial surface density profile that follows a power law with $n=1.5$ (Eq. \ref{eq:Rtrunc_param}). Hence in this regime $\Delta M \propto \rhoism^{-13/35} \vism^{-33/35}$, i.e. the amount of accreted ISM decreases with both increasing density and velocity. The decrease depends more strongly on the velocity than on the density, as can also be seen in Fig. \ref{fig:Mend_param}. For $\vism \lesssim 2.7$ km/s Bondi-Hoyle accretion dominates the accretion process, so at these velocities our model does not provide a good estimate for the amount of ISM that can potentially be accreted. Here we only want to single out the effects of face-on accretion. 

Figs. \ref{fig:Rend_param} illustrates that protoplanetary disc radii can be significantly reduced if they reside in a dense gaseous environment for a long time. Assuming the disc has a similar initial mass but a larger radius, i.e. a lower surface density, would make the process more severe. In the hypothetical case where accretion continues for 10 Myr, the final disc radii are reduced by an additional factor of about 0.4 compared to the contours shown in Fig. \ref{fig:Rend_param}, and $\Delta M$ increases only by about 60\,\% with respect to Fig. \ref{fig:Mend_param}. Considering the duration and typical gas density of the embedded phase, this implies that it is unlikely that the amount of swept-up ISM exceeds the initial mass of the protoplanetary disc, even if discs accrete from the ISM for several Myr. In future work, we will apply our model in a more physically representative environment to quantify the effects it could have in star-forming regions.

\section{Discussion}\label{sec:dis_param}

We discuss the uncertainties in the simulations and how they can affect the validity and predictions of our theoretical model. Then we consider the effect of the simplifying assumptions made and we end this section with an outlook on the implications of our model.

\subsection{Interstellar medium resolution}\label{sec:dis_resolution_param}

As the SPH particles in all simulations have approximately the same mass (see Sect. \ref{sec:setup_param}), the simulations at the lowest ISM density have the lowest resolution with the ISM flow consisting of roughly 500 SPH particles in the computational domain. As discussed in Sect. \ref{sec:setup_param}, this resolution still corresponds to several thousands of particles that interact with the disc on the timescale of the simulation. However, at this low SPH particle number the density is not very well sampled and it is not distributed homogeneously over the computational domain. When the surface area of the disc is small, the number of ISM particles interacting with the disc are even smaller and the results are subject to increased noise. In simulations V20N4 and V30N4, the cross section of the disc is smallest and therefore these simulations are most sensitive to low resolution. Indeed, Figs. \ref{fig:Mcomp_param} and \ref{fig:Mdotcomp_param} show that the ratio between the simulated and theoretical accretion rate and accreted mass decreases with increasing velocity, if the Bondi-Hoyle regime at $\vism = 1$ km/s is not taken into account, suggesting that this result may indeed depend on the size of the disc, which is smaller for higher velocities. The clumpiness and noise in the simulations with the lowest density suggest that the resolution of SPH particles may be too low to properly model the process of face-on accretion. When the number of SPH particles and, correspondingly the density, is increased by a factor of 10, i.e. in simulations V20N5 and V30N5, the accretion rate does agree with our theoretical model. In these cases the disc surface area decreases by a factor of about 4, so the number of incident SPH particles onto the disc is a factor of 2.5 higher than at the lowest density. Furthermore, in simulation V3N5, the mass flux and SPH particle flux is the same as in simulation V30N4, but in the former simulation the density is better sampled across the computational domain. Although the disc radii are similar, the agreement with our analytical model is better for V3N5 in the sense that we do not underestimate the accretion rate in this case. We cannot rule out that this is caused by gravitational focussing at the lower velocity, but since the disc radius is only slightly smaller than the Bondi-Hoyle radius, this is only a small effect. Therefore, the underestimate of the accretion at the highest velocity and lowest density is likely to be at least partly a numerical effect. Unfortunately, verification of this would require more expensive high resolution simulations in which the necessary time steps in our code would become very small and imply a prohibitively long runtime.

\subsection{Stripping of the disc}\label{sec:dis_stripping_param}

Table \ref{tb:strippeddisc_param} and Fig. \ref{fig:discmass_param} show that, in particular at the lowest density, material is continuously stripped from the disc by the ISM flow. We found the same effect in Paper I, where we studied one case, $\vism = 20$ km/s and $\rhoism = 1.9 \times 10^{-17}$ g/cm$^3$, extensively with two different SPH codes and at different resolutions. The stripped disc material also carried angular momentum away from the disc. We pointed out that there is a decreasing trend with increasing resolution in our simulations, suggesting that this mass and angular momentum loss is caused by a numerical artefact (see Sect. 5.1 of Paper I). The simulation with the same parameters in this work, V20N6, confirms this decreasing trend with resolution. The resolution we used in this work is eight times higher than the highest resolution of the simulations in Paper I using the same SPH code. For model V20N6, we calculated the mass and angular momentum loss rates, $\dot{M}_{\rm strip}$ and $\dot{J}_{\rm strip}$, due to continuous stripping of disc material over the same time interval as in Paper I, i.e. between 1500 and 2500 year. Both rates have decreased by almost a factor of 2 with respect to the values in Table 4 of Paper I. In other work, i.e. \citet{moeckel09}, it was found that the mass lost from the disc could be completely accounted for by accretion onto the star, as discussed in Paper I. 

The stripping that is not predicted by our theoretical model is worst at the lowest density, i.e. the lowest resolution of the ISM. It is likely that at this resolution a single SPH particle has a higher (numerical) impact on the disc than it would have if the density was sampled by a higher number of SPH particles. In nature, the density is not perfectly homogeneous either but for a fair comparison with our theoretical model it should be homogeneous in our simulations. Even though some stripping of mass and angular momentum from the disc occurs in our simulations, the evolution of the disc radius and accretion onto the disc is consistent with our theoretical model. We therefore argue that our model describes the process of face-on accretion adequately and can be used as a conservative estimate for the disc radius and the amount of accreted ISM onto the protoplanetary disc within the appropriate timescale (Eq. \ref{eq:tauvisc-dom_param}).

\subsection{Effects of ISM turbulence}

We have assumed that the ISM flow has no angular momentum, but in principle the ISM could also carry angular momentum in turbulent substructures. In order to effectively add this angular momentum to the disc, these substructures must be comparable or larger in size than the radius of the disc. There is no reason to assume that such turbulence in the ISM would have a preferred orientation. If the disc encounters a region of the ISM in which the angular momentum would add to that of the disc, then it is equally likely to encounter a region with an adverse configuration that decreases its angular momentum at a later time. For individual discs, we expect a random-walk process in which some discs gain net angular momentum and some discs lose angular momentum. The magnitude of this effect scales as $\sqrt{\Delta M_{\rm ISM}}$. However, when studying the effect on a large sample of discs the average is zero. Therefore, although this assumption does not apply to individual discs, our model is still applicable for a population study of accretion discs.

\subsection{Viscous evolution of the disc}

The applicability of our analytical model for face-on accretion is limited by our neglect of viscous effects. In Sect. \ref{sec:timescales_param} we estimated a timescale on which viscous effects become important throughout the whole disc.  This timescale is between $4\times 10^4$ and $2\times 10^5$ yr for the velocities and densities adopted in our simulations. On longer timescales, viscous effects can no longer be ignored. To assess these effects in a qualitative manner, we can make use of Eqs.~\ref{eq:dr-dt-timescales_param} and \ref{eq:dSigma-dt-timescales}. As we saw in Sect. \ref{sec:model_param}, ISM accretion tends to flatten the surface density profile over time, which becomes independent of the radius at very large $\tau$. Denoting by $p$ the local slope of the surface density profile, $p = -\partial\log\Sigma(r,t)/\partial\log r$, this means $p < n$ for $\tau \gtrsim 1$ and $p \rightarrow 0$ for $\tau \gg 1$. Taking $n = \frac{3}{2}$ (as appropriate for an isothermal disc), we see that the viscous terms tend to shrink the disc further ($\der r/\der t < 0$ in Eq.~\ref{eq:dr-dt-timescales_param}) and increase the surface density; the second term in Eq.~\ref{eq:dSigma-dt-timescales} is positive for $p < \frac{3}{2}$. This viscous term can be written as
\begin{equation}
\bigg( \frac{\partial\Sigma}{\partial t} \bigg)_{\nu} \approx (2-p)({\textstyle\frac{3}{2}}-p)\,\frac{\Sigma_0(r_0)}{\tau_{\nu}(r_0)}\,\bigg(\frac{r}{r_0}\bigg)^{-1/2},
\end{equation}
making use of Eq.~\ref{eq:tauvisc_param}. Thus $\Sigma$ increases faster at smaller radii than at larger radii, i.e.\ the slope $p$ of the surface density profile tends to increase again. The viscous term thus attempts to restore the $r^{-3/2}$ profile, while continued accretion tends to flatten it. In conclusion, the main effect of viscosity is to speed up the overall contraction of the disc caused by ISM accretion. 

However, the effect on the outer disc radius is more subtle. When subject to ISM accretion, the enhanced overall contraction of the disc that results from viscosity must be accompanied by some outward spreading of the edge of the disc to ensure angular momentum conservation. This tends to increase the outer disc radius in comparison to the inviscid case described by our analytical model. On the other hand, in the regime of high velocities and densities, outward spreading of the disc also makes it more vulnerable to continued ram-pressure stripping. To some extent, these counteracting effects balance each other, such that viscous processes probably do not strongly affect the face-on accretion rate. Our simulations show that despite the presence of viscous spreading at the inner and outer edge of the disc, our analytical model can still be used as a conservative estimate for the disc radius and accretion rate within the viscous timescale, although perhaps not on longer timescales.

When the disc moves from a region of high to low density, viscous spreading is expected to increase the size of the disc and, hence, change the accretion rate. This can be incorporated in the theoretical framework, which we provide but is beyond the scope of this work.

\subsection{Inclination}\label{sec:dis_alignment_param}

We have assumed that the disc is aligned perpendicularly to the flow, which is the most favourable orientation for face-on accretion and provides a straightforward way to test our theoretical model. In reality we may expect initially random orientations, so that for a typical disc the mass flux is lower and the quantitative effects may be smaller than we describe in this paper. In our analytical model (Eq. \ref{eq:dr-dt-viscous_param} and Eq. \ref{eq:dSigma-dt-viscous_param}) the inclination $i$ between the disc axis and the flow direction can be simply accounted for by taking the velocity component perpendicular to the disc plane, i.e. by replacing $\vism$ by $\vism \cos i$. However, the inclination angle angle itself may also change as a result of the flow and accretion process. In a future work (Wijnen et al., in preparation) we investigate the consequences of a non-zero inclination between the disc and flow.

\subsection{Consequences for planet formation}

Our model indicates that in star-forming regions with a high gas density, the protoplanetary discs are expected to be more compact and have a higher surface density when compared to star-forming regions that are less dense. The current consensus is that hot Jupiters, i.e. Jupiter-mass planets with orbital periods $\lesssim$ 10 days, cannot have formed in situ but must have formed at larger radii where more disc material is present and temperatures are lower (e.g. \citealt{lin96}, although \citealt{batygin16} discuss an in-situ formation scenario). On the other hand, warm Jupiters, similar planets with orbits of 10-200 days, are thought to fall in two categories likely reflecting migration and in-situ formation \citep{huang16}. As the process of face-on accretion causes disc material to migrate inwards, it may aid in getting the material in place to form these planets in situ or aid in their inward migration. 
To what extent this process plays a role in the formation and/or migration of hot and warm Jupiters has to be tested by incorporating our findings into detailed planet formation and migration models, which also take into account the appropriate viscous timescales at these radii.

In general, our model suggests that more compact planetary systems may be formed in dense, (formerly) gaseous environments compared to tenuous conditions or environments. A population study of the occurrence of hot and warm Jupiters \citep[possibly with the help of Gaia;][]{dzigan14} in different environments might provide more insight on whether the compactness of a planetary system depends on its birth environment.

\section{Conclusions}\label{sec:conclusion_param}

We have presented a theoretical framework to estimate the radii of and accretion onto protoplanetary discs as a consequence of their face-on movement through an ambient medium. We neglected viscous effects in the disc in this work, but they can be incorporated into our model. We also found a prescription for the radius to which the disc is truncated as an effect of ram pressure stripping that is more restrictive than that currently used in the literature. We tested our model by performing smoothed particle hydrodynamics simulations for a range of velocities (1, 3, 5, 10, 20 and 30 km/s) and densities ($1.9 \times 10^{-19}$, $1.9 \times 10^{-18}$ and $1.9 \times 10^{-17}$ g/cm$^3$). We find a good agreement between the simulated and theoretically predicted evolution of the disc radii and ISM accretion after the simulations reached a steady state. In general, our model tends to slightly underestimate the disc radii and hence the accretion rate onto the protoplanetary discs compared to the simulations. We find that the effective cross section of the disc is smaller than the actual surface area because part of the ISM is deflected around the outer edge of the disc. This confirms the findings of previous work, for example \citet{ouellette07} and \citet{wijnen16}. As our theoretical model tends to underestimate the radius of the disc compared to the simulations, the predicted radius provides a better estimate for the effective cross section of the disc and hence for the accretion rate. Our theoretical framework can therefore be used as a conservative estimate for these quantities. The differences that we find between our model and the simulations arise in part from the simplicity of our theoretical model, but they can also be partly ascribed to numerical effects in the simulations.

In our simulations, the disc loses mass due to continuous stripping by the ISM flow. This mass loss is not taken into account in our theoretical model and we cannot determine if, and to what extent, this stripping is numerical or physical. We found in previous work \citep{wijnen16} that the stripping rate decreases with increasing resolution and this trend is confirmed in this work. Other works, for example \citet{moeckel09}, do not find any significant stripping of the disc. The stripping does not appear to have a significant influence on the evolution of the disc radii and ISM accretion as described by our model and as seen in the simulations.

Our theoretical model also correctly captures the main features of the evolution of the surface density profile in the simulations, but there are quantitative differences. These are partly caused by the surface density profile in our model having a sharp cut-off at the inner and outer edge of the disc, while this is not the case in the simulations and in reality. In addition, the (numerical) viscous processes in the disc also play a role in the redistribution of material in the disc, which we have not taken into account in the comparison with our theoretical model. Nonetheless, our model describes the evolution of the radius and accretion rate well on a timescale of $10^4$ yr, which is the duration of the simulations. 

Our model predicts that the radii of protoplanetary discs can be severely reduced in a dense gaseous environment. The impact of ram pressure stripping is most severe for velocities $>$ 10 km/s and densities $\gtrsim 10^{-18}$ g/cm$^3$, which may be expected in massive star-forming regions and around winds from supergiants and interacting binaries; even higher relative velocities can be expected in interactions with supernova ejecta or winds from massive stars. The continuous accretion of ISM at lower velocities and densities can also substantially decrease the radii of protoplanetary discs. For a typical velocity of 3 km/s and density of $2 \times 10^{-19}$ g/cm$^3$ in the core of a star-forming region (see Sects. \ref{sec:densities_param} and \ref{sec:velocities_param}) the radius is halved, from 100 to 50 AU, and the disc has accreted 25\,\% of its initial mass after 1 Myr. Both processes are relatively insensitive to the timescale $\tau$ as the radius scales as $\tau^{-2/5}$ and the amount of accreted material as $\tau^{1/5}$. Owing to this accretion process, the discs become more compact, as material migrates to smaller radii, potentially making it more likely to form (massive) planets close to the star. Hence our model suggests that planetary systems can be more compact in (formerly) dense star-forming regions than in more tenuous regions and that the occurrence of hot Jupiters may depend on the birth environment of their host stars. 

Future studies should provide more insight into the conditions at which different disc truncation processes are relevant. For example, it may be expected that at high stellar densities the effect of close encounters outweighs the high relative velocity, which is necessary for ram pressure stripping and contraction to be effective. Likewise, photo-evaporation may only become important at later stages of embedded star-forming regions, when the ambient gas density is low and protoplanetary discs still have radii that are sufficiently large to be affected by photo-evaporation.

\begin{acknowledgements}
We are grateful to Carlo Manara, Giovanni Rosotti, Stefano Facchini, Nate Bastian, Sebastian Ohlmann, Tim Lichtenberg, Matthew Kenworthy, Carsten Dominik, and Hilke Schlichting for  valuable discussions. We thank the referee for his/her comments. This research is funded by the Netherlands Organisation for Scientific Research (NWO) under grant 614.001.202.
\end{acknowledgements}

\bibliographystyle{aa} 
\bibliography{phdbib}

\begin{appendix}
\section{Equations for disc evolution}
\subsection{Derivation}\label{ap:disc_eqs_deriv}

We derive the equations governing accretion disc evolution, following Chapter 5.2 of \citet[ hereafter FKR02]{frank02}. We modify the equations to allow for accretion of gas from the ISM, with density $\rhoism$ and velocity $\vism$, perpendicular to the plane of the disc. The surface density of the disc at time $t$ and radius $r$ is $\Sigma(r,t)$. Conservation of mass gives
\begin{equation} \label{eq:mass-cons_param}
r \frac{\partial\Sigma}{\partial t} = \rhoism \vism \, r - \frac{\partial}{\partial r} (r \Sigma v_r)
\end{equation}
(compare to Eq.~5.3 of FKR02). Here $v_r \equiv \der r/\der t$ is the radial velocity of matter within the disc, i.e.\ the Lagrangian derivative of the radius. The partial derivatives are taken at constant $r$ and $t$. Because the ISM gas carries no angular momentum, the angular-momentum conservation equation remains as given by Eq.~5.4 of FKR02,
\begin{equation} \label{eq:AM-cons_param}
r \frac{\partial}{\partial t} (\Sigma\, j) + \frac{\partial}{\partial r} (r \Sigma v_r\, j) = \frac{1}{2\pi}\, \frac{\partial G}{\partial r},
\end{equation}
where $j(r) = r^2\Omega(r)$ is the specific angular momentum of the disc material at radius $r$ and $\Omega(r)$ is the angular velocity. The function $G(r,t)$ describes the viscous torques operating within the disc. Combining both equations above, together with the assumption that $\partial j(r)/\partial t = 0$, gives
\begin{equation} \label{eq:combined_param}
j \, \rhoism \vism \, r + r \Sigma v_r \frac{\partial j}{\partial r} = \frac{1}{2\pi}\, \frac{\partial G}{\partial r},
\end{equation}
which is the equivalent of Eq.~5.6 of FKR02. This can be rewritten as an explicit expression for $r \Sigma v_r$ and substituted into Eq.~\ref{eq:mass-cons_param} to give
\begin{eqnarray}
\frac{\partial\Sigma}{\partial t} & = &\rhoism \vism \\
&&- \frac{1}{r} \frac{\partial}{\partial r} \Bigg[ \frac{1}{2\pi\, \partial j/\partial r} \frac{\partial G}{\partial r} - \rhoism \vism \, r \frac{j}{\partial j/\partial r} \Bigg], \nonumber
\end{eqnarray}
For a Keplerian disc, where the mass $M$ of the central object is much higher than the disc mass \Mdisc, we have $j(r) = \sqrt{GMr}$ and $\partial j/\partial r = \sqrt{GMr}/2r$. In this case the second term in the square brackets yields a contribution $4 \rhoism \vism$, and we obtain
\begin{equation}
\frac{\partial\Sigma}{\partial t} = 5 \rhoism \vism - \frac{1}{r} \frac{\partial}{\partial r} \Bigg[ \frac{r^{1/2}}{\pi \sqrt{GM}} \frac{\partial G}{\partial r} \Bigg].
\end{equation}
Using the relation for $G(r,t)$ given by Eq.~5.5 of FKR02, i.e.\
\begin{equation}
G(r,t) = 2\pi r \, \nu\Sigma\, r^2\frac{\partial\Omega}{\partial r}, 
\end{equation}
we obtain a differential equation involving the viscosity $\nu$,
\begin{equation} \label{eq:dSigma-dt-viscous_app_param}
\frac{\partial\Sigma}{\partial t} = 5 \rhoism \vism + \frac{3}{r} \frac{\partial}{\partial r} \Bigg[ r^{1/2} \frac{\partial}{\partial r} (r^{1/2} \nu\Sigma) \Bigg].
\end{equation}
Finally we get an expression for the radial drift velocity $v_r$ from Eq.~\ref{eq:combined_param},
\begin{equation} \label{eq:dr-dt-viscous_app_param}
v_r = - 2\rhoism \vism \, \frac{r}{\Sigma} - \frac{3}{r^{1/2}\Sigma} \frac{\partial}{\partial r} (r^{1/2} \nu\Sigma).
\end{equation}
Equations~\ref{eq:dSigma-dt-viscous_app_param} and \ref{eq:dr-dt-viscous_app_param} are equivalent to Eqs.~5.8 and 5.9 in FKR02, with the addition of an ISM accretion term involving $\rhoism \vism$ in each equation.

\subsection{Viscous effects}\label{app:viscous_effects_param}

Although viscous effects can be neglected at least in the outer parts of the disc and at early times, this is no longer justified over very long timescales. To assess the importance of disc viscosity as a function of time, we compare the viscous and accretion timescales near the outer edge of the disc, where the effect of ISM accretion dominates over viscous effects for the longest time (Sect. \ref{sec:timescales_param}). Let us assume that the disc manages to remain isothermal at constant temperature (i.e.\ the disc is always able to radiate efficiently), so that $c_s^2$ is a constant, and for simplicity we assume that $\rhoism \vism$ is constant. Applying Eqs.~\ref{eq:tauvisc_param} and \ref{eq:taum_param} from Sect. \ref{sec:timescales_param} for $r = \Rout$ and taking $r_0 = R_\mathrm{out,0}$ we obtain
\begin{equation} \label{eq:tvisc-tau_param}
\tau_{\nu}(\Rout) = \frac{\sqrt{GMr_0}}{3\alpha c_s^2}\, \yout^{1/2} ~ \approx ~ \tau_{\nu}(r_0)\cdot \tau^{-1/5} \quad \mbox{for} \quad \tau \gg 1,
\end{equation}
\begin{equation} \label{eq:tacc-tau_param}
\tau_{\dot{m}}(\Rout) = \frac{\Sigma_0(r_0)}{5\rhoism \vism}\,(\yout^{-n} + \tau) ~ \approx ~ \tau_{\dot{m}}(r_0) \cdot\tau \quad \mbox{for} \quad \tau \gg 1,
\end{equation}
Where $\tau$ is defined by Eq. \ref{eq:tau-integral}. For a constant mass flux, $\tau = t/\tau_{\dot{m}}(r_0)$. We suppose $\tau_{\nu}(r_0) > \tau_{\dot{m}}(r_0)$, i.e.\ the outermost parts of the disc are initially accretion dominated. This remains the case as long as 
\begin{equation} \label{eq:tau-limit_param}
\tau \lesssim \bigg[ \frac{\tau_{\nu}(r_0)}{\tau_{\dot{m}}(r_0)} \bigg]^{5/6}, \quad \mbox{or} \quad t \lesssim \tau_{\nu}(r_0)^{5/6}\tau_{\dot{m}}(r_0)^{1/6}
.\end{equation}
On longer timescales, viscous effects are important over the entire disc.

\subsection{Gravitational instability}\label{sec:grav_instability_param}

Gaseous discs that are differentially rotating can become gravitationally unstable if the Toomre parameter $Q$ \citep{toomre64} is less than unity at the outer edge of the disc,
\begin{equation}\label{eq:ToomreQ_param}
Q = \frac{c_s \Omega}{\pi G \Sigma}
.\end{equation}
As the surface density profile of the disc increases because of accretion and contraction, the disc in our model could in theory become unstable against gravitational instabilities. We can solve Eq. \ref{eq:ToomreQ_param}, assuming $Q=1$, to derive an expression for the critical surface density profile. We can then equate this expression to Eq. \ref{eq:sigmari_param} to determine on what timescale gravitational instabilities become important. This timescale is proportional to, among other parameters that do not vary between our simulations, $(\rhoism \vism)^{-1}$, and is therefore the shortest for the highest mass flux. For our simulations, that is when $\vism = 30\,$km/s and $\rhoism = 1.9 \times 10^{-17}$g/cm$^3$, which corresponds to a timescale of $2 \times 10^4$ yr at 20 AU, i.e. roughly the disc radius at that time. This is of the same order as  $\tau_{\nu,\,\mathrm{dom}}$ of $4 \times 10^4$ yr in that case (see Sect. \ref{sec:timescales_param}). However, typically this timescale is higher than $\tau_{\nu,\,\mathrm{dom}}$, for example for the case shown in Fig. \ref{fig:timescales_param} it is $2 \times 10^6$ yr at 90 AU compared to $\tau_{\nu,\,\mathrm{dom}}$ of $1.7 \times 10^5$ yr. We conclude that the disc does not become gravitational unstable within the timescale that our model can be used to describe the evolution of the disc.

\end{appendix}

\end{document}